\documentclass[twocolumn,showpacs,preprintnumbers,nofootinbib,prd,superscriptaddress,10pt]{revtex4-1}
\usepackage{graphicx,amssymb,amsmath,amsthm,amsfonts,epsfig}

\usepackage[linktocpage]{hyperref}
\usepackage[usenames,dvipsnames]{color}
\usepackage{epstopdf}
\usepackage{aas_macros}
\definecolor{darkred}{rgb}{0.5,0,0}
\definecolor{darkgreen}{rgb}{0,0.5,0}
\definecolor{darkblue}{rgb}{0,0,0.5}
\definecolor{prussian}{rgb}{0.0, 0.19, 0.33}
\definecolor{richelectricblue}{rgb}{0.03, 0.57, 0.82}
\definecolor{mediumseagreen}{rgb}{0.24, 0.7, 0.44}
\definecolor{lust}{rgb}{0.9, 0.13, 0.13}
\definecolor{ballblue}{rgb}{0.13, 0.67, 0.8}
\definecolor{darkcyan}{rgb}{0.0, 0.55, 0.55}
\definecolor{mountainmeadow}{rgb}{0.19, 0.73, 0.56}
\definecolor{palecarmine}{rgb}{0.69, 0.25, 0.21}
\definecolor{richcarmine}{rgb}{0.84, 0.0, 0.25}
\definecolor{tangelo}{rgb}{0.98, 0.3, 0.0}
\definecolor{venetian}{rgb}{0.784,0.031,0.082}
\definecolor{bdfrance}{rgb}{0.192,0.549,0.906}

\hypersetup{colorlinks=true, citecolor=venetian,
linkcolor=bdfrance, urlcolor=lust}
\usepackage{amsmath,amssymb}
\usepackage{tensor}
\usepackage{mathtools}
\usepackage{amsbsy}
\usepackage{bm}
\usepackage{float}

\newcommand{\be}{\begin{equation}}
\newcommand{\ee}{\end{equation}}
\newcommand{\bear}{\begin{eqnarray}}
\newcommand{\eear}{\end{eqnarray}}

\newcommand{\p}{\prime}

\newcommand{\cE}{{\cal E}}

\newcommand{\cC}{{\cal C}}

\newcommand{\cP}{{\cal P}}
\newcommand{\cM}{{\cal M}}

\begin{document}
\title{Post-Tolman-Oppenheimer-Volkoff formalism for relativistic stars}

\author{Kostas Glampedakis}
\email{kostas@um.es}
\affiliation{Departamento de F\'isica, Universidad de Murcia,
Murcia, E-30100, Spain}
\affiliation{Theoretical Astrophysics, University of T\"ubingen, Auf der Morgenstelle 10, T\"ubingen, D-72076, Germany}

\author{George Pappas}
\email{georgios.pappas@nottingham.ac.uk}
\affiliation{School of Mathematical Sciences, The University of Nottingham,
University Park, Nottingham NG7 2RD, United Kingdom}

\author{Hector O. Silva}
\email{hosilva@phy.olemiss.edu}
\affiliation{Department of Physics and Astronomy,
The University of Mississippi, University, MS 38677, USA}

\author{Emanuele Berti}
\email{eberti@olemiss.edu}
\affiliation{Department of Physics and Astronomy,
The University of Mississippi, University, MS 38677, USA}
\affiliation{CENTRA, Departamento de F\'isica, Instituto Superior T\'ecnico, Universidade de Lisboa, Avenida Rovisco Pais 1, 1049 Lisboa, Portugal}

\date{{\today}}

\begin{abstract}

  Besides their astrophysical interest, compact stars also provide an
  arena for understanding the properties of theories of gravity that
  differ from Einstein's general relativity. Numerous studies have
  shown that different modified theories of gravity can modify the
  bulk properties (such as mass and radius) of neutron stars for given
  assumptions on the microphysics. What is not usually stressed though
  is the strong degeneracy in the predictions of these theories for
  the stellar mass and radius.  Motivated by this observation, in this
  paper we take an alternative route and construct a stellar structure
  formalism which, without adhering to any particular theory of
  gravity, describes in a simple parametrized form the departure from
  compact stars in general relativity.  This
  ``post-Tolman-Oppenheimer-Volkoff (TOV)'' formalism for spherical
  static stars is inspired by the well-known parametrized
  post-Newtonian theory, extended to second post-Newtonian order by
  adding suitable correction terms to the fully relativistic TOV
  equations.  We show how neutron star properties are modified within
  our formalism, paying special attention to the effect of each
  correction term.  We also show that the formalism is equivalent to
  general relativity with an ``effective'' (gravity-modified) equation
  of state.
\end{abstract}


\maketitle



\section{Introduction}
\label{sec:intro}

Neutron stars play a special role among astrophysical objects, because
they are excellent laboratories for matter under extreme conditions
(unlike black holes) and {\em also} excellent laboratories to probe
strong gravity (unlike ordinary stars or white
dwarfs)~\cite{Berti:2015itd}.
For these reasons neutron stars are among the main targets of future
observatories, such as SKA~\cite{Watts:2014tja},
NICER~\cite{2014SPIE.9144E..20A}, LOFT~\cite{Feroci:2011jc} and
AXTAR~\cite{Ray:2010uq}. These experiments have the potential to
measure neutron star masses and radii to unprecedented levels
\cite{Psaltis:2013fha,Miller:2014mca,Miller:2014aaa}. If general
relativity (GR) is assumed to be the correct theory of gravity, the
observed mass-radius relation will constrain the equation of state
(EOS) of matter at supranuclear densities, which is inaccessible to
laboratory
experiments~\cite{Lattimer:2004pg,Lattimer:2006xb,Ozel:2010fw,Steiner:2010fz,Hebeler:2013nza,Miller:2013tca}. A
procedure to reconstruct the EOS from observations of the mass-radius
relation (working within GR) was developed in a series of papers by
Lindblom and collaborators
\cite{1992ApJ...398..569L,Lindblom:2012zi,Lindblom:2013kra};
see~\cite{Lindblom:2014sha} for a review.


\begin{figure}[hbt]
\includegraphics[width=0.5\textwidth]{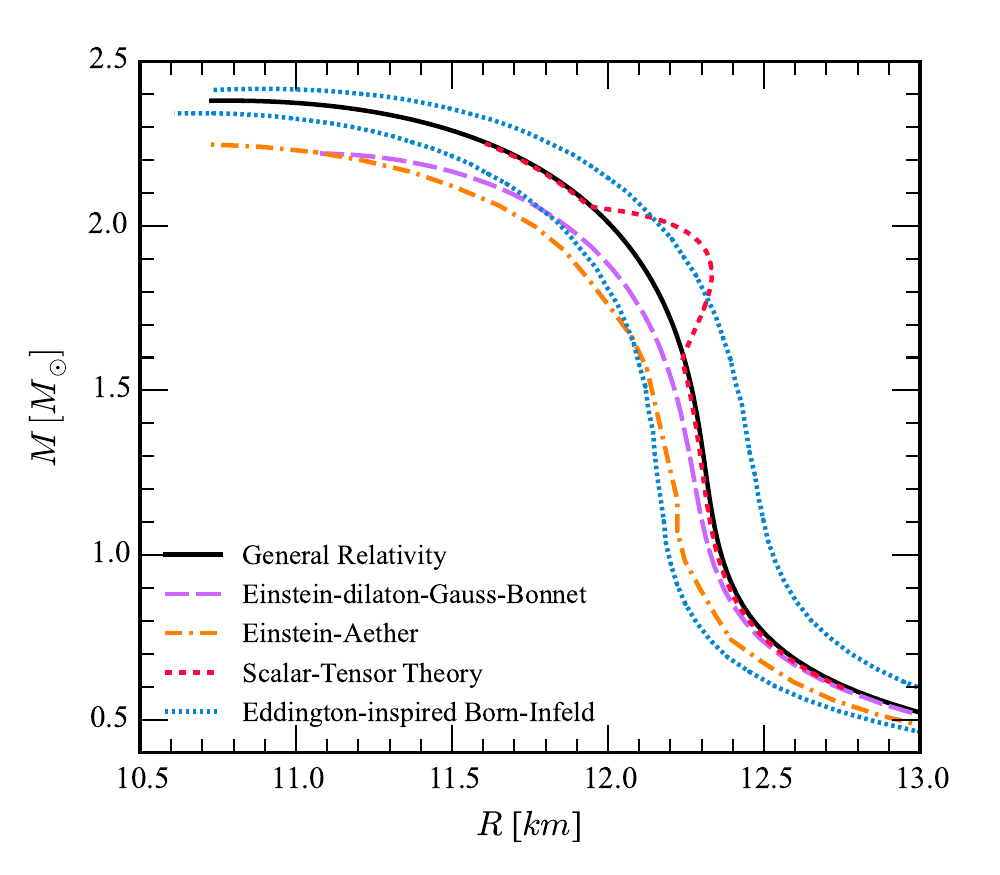}
\caption{{\it The gravity-theory degeneracy problem.} The mass-radius 
  relations in different modified theories of gravity for EOS 
  APR~\cite{Akmal:1998cf}. Masses are measured in solar masses, and 
  radii in kilometers. The theory parameters used for this plot are:
  $\alpha = 20\,M^2_{\odot}$ and $\beta^2 = 1$
  (Einstein-dilaton-Gauss-Bonnet~\cite{Pani:2011xm}); $c_{14} = 0.3$
  (Einstein-Aether~\cite{Yagi:2013ava}); $\beta = -4.5$ (scalar-tensor 
  theory~\cite{Pani:2014jra}) and $\kappa = \pm 0.005$
  (Eddington-inspired-Born-Infeld gravity~\cite{Sotani:2014goa}). 
  Even if the high-density EOS were known, it would be hard to 
  distinguish the effects of competing theories of gravity on the bulk 
  properties of neutron stars.}
\label{fig:theory_deg}
\end{figure}

Besides their interest for nuclear physics, neutron stars are also unique probes
of strong-field gravitational physics. For any given EOS, theories
that modify the strong-field dynamics of GR generally predict bulk
observable properties (neutron star mass, radius, moment of inertia and higher
multipole moments) that are different from those in Einstein's theory.
However, a survey of the literature on neutron stars in modified
theories of gravity (see e.g. Table~3 of~\cite{Berti:2015itd}) reveals
a high degree of degeneracy in the salient properties of relativistic
stars.
As we show in Fig.~\ref{fig:theory_deg}, if we assume a
nuclear-physics motivated EOS (specifically, EOS
APR~\cite{Akmal:1998cf} in the figure), modifications in the gravity
sector are usually equivalent to systematic shifts of the GR
mass-radius curves towards either higher masses and larger radii (as
in the case of scalar-tensor
theories~\cite{Pani:2014jra,Doneva:2013qva}), lower masses and smaller
radii (as in the case of
Einstein-dilaton-Gauss-Bonnet~\cite{Pani:2011xm,Kleihaus:2014lba}
and Lorentz-violating theories~\cite{Yagi:2013qpa,Yagi:2013ava}) or
both, as in Eddington-inspired-Born-Infeld gravity with different
signs of the coupling parameter~\cite{Pani:2012qb,Sotani:2014goa}.

Systematic shifts in the mass-radius relation could be attributed
either to the poorly known physics controlling the high-density EOS,
or to modifications in the theory of gravity itself. This EOS/gravity
degeneracy is intrinsic in all attempts to constrain strong gravity
through astrophysical observations of neutron stars: chapter 4
of~\cite{Berti:2015itd} reviews various proposals to solve this
problem, e.g. through the recently discovered universal relations
between the bulk properties of neutron
stars~\cite{Yagi:2013bca,Yagi:2013awa,Pappas:2013naa,Yagi:2014bxa}.

In any case, different gravitational theories span (at least
qualitatively) the same parameter space in terms of their predictions
for relativistic stellar models. Gravity-induced modifications usually
look like smooth deformations of the general relativistic
predictions. A notable exception are cases where nonperturbative
effects induce phase transitions, as in the ``spontaneous
scalarization'' scenario first proposed in~\cite{Damour:1993hw}, where
modifications only occur in a specific range for the central density.

With the possible exception of nonperturbative phase transitions,
these considerations suggest that the broad features of a large class
of modified gravity theories can be reproduced, at least for small
deviations from GR, by a perturbative expansion around a background
solution given by the standard TOV equations, which determine the
structure of relativistic stellar models in
GR~\cite{Misner1973,poisson2014gravity}.

Instead of committing to one particular pet theory, in this paper we
formulate a parametrized ``post-TOV'' framework for relativistic stars
based on the well-known parametrized post-Newtonian (PPN) theory
developed by Nordtvedt and Will \cite{Will:1972zz,Nordtvedt:1972zz};
see e.g.~\cite{Will:1993ns,poisson2014gravity} for introductions to
the formalism.
The foundations of post-Newtonian (PN) theory for fluid configurations
in GR were laid in classic work by Chandrasekhar and
collaborators~\cite{1965ApJ...142.1488C,1969ApJ...158...55C}. Various
authors studied stellar structure using the PN approximation, both in
GR
\cite{Asada:1996ai,Taniguchi:1998wi,Shinkai:1998mg,Gupta:2000zb} and
in modified theories of gravity, such as scalar-tensor theory
\cite{Nutku:1969,Xie:2007gq}.  To our knowledge, after some early work
that will be discussed
below~\cite{WagonerMalone:1974,CiufoliniRuffini:1983,1976ApJ...207..263S},
the investigation of compact stars within the PPN approximation has
remained dormant for more than thirty years. In the intervening time
the PPN parameters have been extremely well constrained by Solar
System and binary pulsar observations at 1PN order
(see~\cite{Will:2014xja} for a review of current bounds).

In this paper we build a phenomenological post-TOV framework by
considering 1PN and 2PN order corrections to the TOV equations. Our strategy
is, at heart, quite simple: from a suitable set of PPN hydrostatic equilibrium equations
we isolate the purely non-GR pieces. These PPN terms are subsequently added ``by hand''
to the {\em full} general relativistic TOV equations, hence producing a
set of parametrized post-TOV equations (cf.~\cite{Yunes:2009ke} for a
similar ``post-Einsteinian'' parametrization in the context of
gravitational radiation from binary systems).
The formalism introduces a new set of 2PN parameters that are
presently unconstrained by weak-field experiments, and that encompass
the dominant corrections to the bulk properties of neutron stars in GR
in a wide class of modified gravity theories.


\begin{figure*}[tbh]
\includegraphics[width=\textwidth]{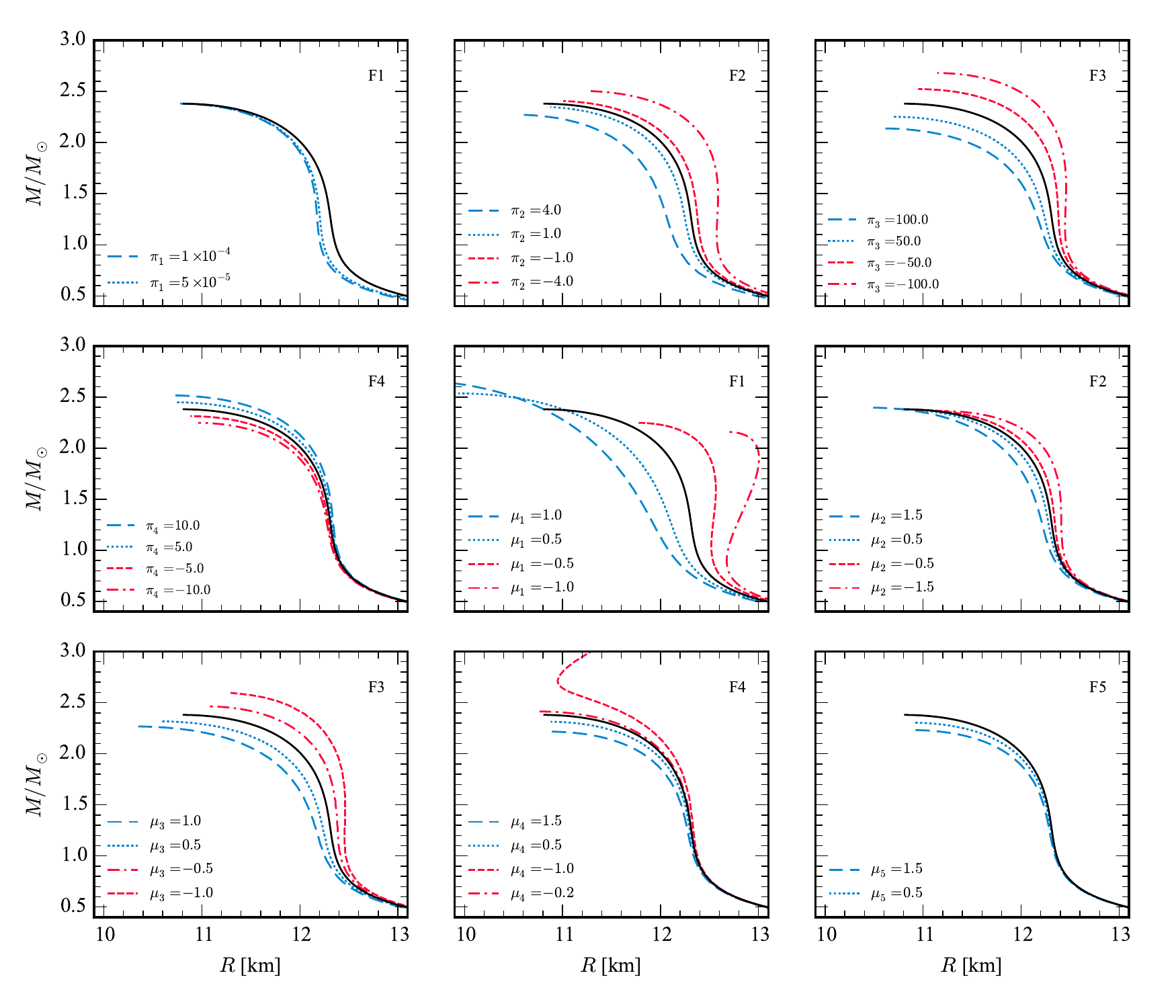}
\caption{{\it 2PN-order post-TOV corrections on the mass-radius
    curves.} We show the modification induced by different families of
  post-TOV terms on the general relativistic mass-radius curve, assuming the APR EOS.
  Left to right and top to bottom, the different panels show the effect of
  the pressure terms, proportional to $\pi_i$ ($i=1,\dots,4$), and of
  the mass terms, proportional to $\mu_i$ ($i=1,\dots,5$).}
\label{fig:MR_families}
\end{figure*}

\subsection{Executive summary}
\label{sec:summary}

Since this paper is rather technical, we summarize our main
conclusions here. The core of our proposal is to use the following set
of ``post-TOV" equations of  structure for spherically symmetric stars (from now on
we use geometrical units $G=c=1$):
\begin{subequations}
\label{PTOV_2PN_intro}
\begin{align}
\frac{dp}{dr} &= \left (\frac{dp}{dr} \right )_{\rm  GR}  -\frac{\rho 
                m}{r^2} \left (\, {\cal P}_1 + {\cal P}_2\, \right ),
\label{eq:PTOV_2PN2_dp}
\\
\nonumber \\
\frac{dm}{dr} & = \left ( \frac{dm}{dr} \right )_{\rm GR} + 4\pi r^2
                \rho \left ( {\cal M}_1 + {\cal M}_2\right ),
\label{eq:PTOV_2PN2_dm}
\end{align}
\end{subequations}
where
\begin{subequations}
\label{PandM}
\bear
&&{\cal P}_1 \equiv \delta_1 \frac{m}{r} + 4\pi \delta_2  \frac{r^3 p}{m},
\\
&&{\cal M}_1 \equiv   \delta_3 \frac{m}{r} + \delta_4 \Pi,
\\
&&{\cal P}_2\equiv \pi_1 \frac{m^3}{r^5\rho} + \pi_2 \frac{m^2}{r^2}
+ ~ \pi_3 r^2 p + \pi_4 \frac{\Pi  p}{\rho},
\\
&&{\cal M}_2\equiv \mu_1 \frac{m^3}{r^5\rho}   + \mu_2 \frac{m^2}{r^2}
 +~ \mu_3 r^2 p
\nonumber
+~ \mu_4 \frac{\Pi  p}{\rho} + \mu_5 \Pi^3 \frac{r}{m} .\\
\eear
\end{subequations}
Here $r$ is the circumferential radius, $m$ is the mass function, $p$
is the fluid pressure, $\rho$ is the baryonic rest mass density,
$\epsilon$ is the total energy density, and
$\Pi \equiv (\epsilon-\rho)/\rho$ is the internal energy per unit
baryonic mass. A ``GR'' subscript denotes the standard TOV equations
in GR [cf. Eq.~\eqref{TOV} below, where we appended a subscript ``T''
to the mass function for reasons that will become apparent later];
$\delta_i, \pi_i$ ($i=1,\,\dots \,,4$) and $\mu_i$ ($i=1, \dots\,, 5$)
are phenomenological post-TOV parameters.  The GR limit of the
formalism corresponds to setting all of these parameters to zero, i.e.
$\delta_i, \pi_i, \mu_i \to 0$.

The dimensionless combinations $\cP_1,\cM_1$ and $\cP_2, \cM_2$
represent a parametrized departure from the GR stellar structure and
are linear combinations of 1PN- and 2PN-order terms, respectively.
In particular, the coefficients $\delta_i$ attached to the 1PN terms
are simple algebraic combinations of the traditional PPN parameters:
see Eqs.~(\ref{delta12}) and (\ref{delta34}) below. As such, they are
constrained to be very close to zero by existing Solar System and
binary pulsar observations\footnote{Using the latest constraints on
  the PPN parameters \cite{Will:2014xja} we obtain the following upper
  limits:
  $| \delta_1| \lesssim 6\times 10^{-4}, |\delta_2| \lesssim 7\times
  10^{-3}, | \delta_3 | \lesssim 7\times 10^{-3}, |\delta_4| \lesssim
  10^{-8}$.}:
$|\delta_i| \ll 1$. This result translates to negligibly small 1PN
terms in Eq.~(\ref{PTOV_2PN_intro}): $\cP_1, \cM_1 \ll1$.  On the other
hand, $\pi_i$ and $\mu_i$ are presently unconstrained, and
consequently $\cP_2, \cM_2$ should be viewed as describing the
dominant (significant) departure from GR.

Each of the two combinations $\cP_2$ and $\cM_2$ involves no more than
five dimensionless 2PN terms, but as we show in Section~\ref{sec:pT2}
these terms are representative of five distinct ``families" consisting
of a large number of 2PN terms. Each family is defined by the property
that all of its members lead to approximately self-similar changes in
the stellar mass-radius curves when included in $\cP_2,\cM_2$. In
other words, as we verified by numerical calculations, we can account
for several terms belonging to the same family by taking just one term
from that family (either the dominant one or, when convenient, a much
simpler subdominant one) and varying the corresponding post-TOV
coefficient $\pi_i$ or $\mu_i$.

The qualitative effect of each of the 2PN-order post-TOV terms on the
mass-radius relation is illustrated in Fig.~\ref{fig:MR_families}. The
values of the $\pi_i$ and $\mu_i$ coefficients in each panel of this
figure were chosen for purely illustrative purposes, i.e., we chose
these coefficients to be large enough that they can produce visible
deviations on the scale of the plot.
A first noteworthy feature is that pressure terms {\em typically}
induce corrections that are about an order of magnitude smaller than
mass terms\footnote{A notable exception to this rule is the $\pi_1$
  term, for reasons that will be explained in Section~\ref{sec:num}
  below.}. This can be seen by the larger range of $\pi_i$'s needed to
produce visible changes in the mass-radius curve ($|\pi_2|\leq 4$,
$|\pi_3|\leq 100$ and $|\pi_4|\leq 10$) when compared to the
corresponding corrections in the mass-function equation
($|\mu_2|\leq 1$, $|\mu_3|\leq 1$ and $|\mu_4|\leq 1.5$,
respectively).  Some terms (such as those proportional to $\pi_2$,
$\pi_3$, $\pi_4$, $\mu_3$ and $\mu_5$) induce smooth rigid shifts of
the mass-radius curve, similar to those that would be produced by a
softening or stiffening of the nuclear EOS. Other terms (like those
proportional to $\mu_1$, $\mu_2$ and $\mu_4$) produce more peculiar
features that are more or less localized in a finite range of central
densities. This is interesting, because (for example) it is plausible
to conjecture that some combination of the $\mu_1$ and $\mu_2$
corrections may reproduce the qualitative features of a highly
non-perturbative phenomenon such as spontaneous scalarization, despite
the intrinsically perturbative nature of our formalism.

The punch line here is that each post-TOV correction is qualitatively
different, so we can use the post-TOV formalism as a toolbox to
reproduce the mass-radius curves shown in Fig.~\ref{fig:theory_deg}
for various modified theories of gravity. More ambitiously, it would
be interesting to address the inverse problem, i.e. to find out how
the post-TOV parameters are related to the dominant corrections
induced by each different theory. These issues are beyond the scope of
this paper, but they are obviously crucial to relate our formalism to
experiments, and we plan to address them in future work.

The second main result of this paper has to do with the
``completeness" of our post-TOV formalism, in the sense that the
stellar structure Eqs.~(\ref{PTOV_2PN_intro}) -- if we neglect the
small terms $\cP_1, \cM_1$ -- can be formally derived by a covariantly
conserved perfect fluid stress energy tensor. That is:
\be
\nabla_\nu T^{\mu\nu} = 0, \qquad T^{\mu\nu} = (\epsilon_{\rm eff} + p) u^\mu u^\nu + p g^{\mu\nu},
\ee
where the effective, gravity-modified energy density is
\be
\label{effectiveEOS}
\epsilon_{\rm eff} = \epsilon + \rho {\cal M}_2,
\ee
and the covariant derivative is compatible with the effective post-TOV metric
\be
g_{\mu\nu} = \mbox{diag} [\, e^{\nu(r)}, (1-2m(r)/r)^{-1}, r^2, r^2 \sin^2\theta\,],
\ee
with
\be
\frac{d\nu}{dr} = \frac{2}{r^2\\} \left [\, (1-{\cal M}_2) \frac{m+4\pi r^3 p}{1-2m/r} + m\cP_2 \, \right ].
\ee
Our phenomenological post-TOV formalism is expected to encompass a
large number of alternative theories of gravity, but it is not
completely general, and future extensions may be possible or even
desirable.  As we stated earlier, theories which produce
non-perturbative phase transitions in their stellar structure
equations may not be accurately modeled. The formalism is also
limited by the choice of acceptable 2PN terms out of all dimensionally
possible combinations, based on criteria that have bearing on the
structure of the gravitational field equations (see
Section~\ref{sec:pT2} below).


\subsection{Plan of the paper}

The plan of the paper is as follows. In Section~\ref{sec:PPN} we
introduce the PPN formalism and review previous applications to
relativistic stars (in particular work by Wagoner and Malone
\cite{WagonerMalone:1974} as well as Ciufolini and Ruffini
\cite{CiufoliniRuffini:1983}).
In Section~\ref{sec:pT} we develop the post-TOV formalism to 1PN
order (where all parameters are already constrained to be very close
to their GR values by Solar System and binary pulsar experiments), and
then to 2PN order. We also show the equivalence between the 2PN
post-TOV equations and GR with a gravity-modified EOS
under a minimal set of reasonable assumptions.
In Section~\ref{sec:num} we present some numerical results
illustrating the relative importance of the different post-TOV
corrections.
Some technical material is collected in three appendices.
Appendix~\ref{sec:DA} gives details of the dimensional analysis
arguments used to select the relevant set of 2PN post-TOV
coefficients.
In Appendix~\ref{sec:RLE} we present a brief summary of the relativistic
Lane-Emden theory,  which plays an auxiliary role in the construction of our formalism.
Finally, Appendix~\ref{sec:potentials} shows that certain integral potentials
appearing at 1PN order in the stellar structure equations (namely,
the gravitational potential $U$, the internal energy $E$ and the
gravitational potential energy $\Omega$) can be approximated by linear
combinations of non-integral potentials, so these integral potentials
are ``redundant'' and can be discarded when building our post-TOV
expansion.

\section{Setting the stage: stellar structure within PPN theory}
\label{sec:PPN}

\subsection{The TOV equations}
\label{sec:tov}

A convenient starting point for our analysis is the standard general
relativistic TOV equations, describing hydrostatic equilibrium in
spherical symmetry~\cite{poisson2014gravity}. These are given by the
familiar formulas:
\begin{subequations}
\label{TOV}
\begin{align}
& \left (\frac{dp}{dr} \right )_{\rm GR} = -\frac{(\epsilon + p)}{r^2}  \frac{ (m_{\rm T} + 4\pi r^3 p  )}{  ( 1-2m_{\rm T} /r )},
\label{dpdr_tov}
\\
\nonumber \\
& \left ( \frac{dm_{\rm T}}{dr} \right)_{\rm GR} = 4\pi r^2 \epsilon,
\label{dmdr_tov}
\end{align}
\end{subequations}
where $p$ and $\epsilon$ are the fluid's pressure and energy density,
respectively, and $m_T$ is the mass function (the subscript is used to
distinguish this mass function from similar quantities appearing in
PPN theory, see below).

For later convenience we also write down the 1PN-order expansion of
these equations (for simplicity the subscript ``GR'' is omitted):
\begin{subequations}
\label{TOV1PN}
\begin{align}
& \frac{dp}{dr} = -\frac{m_{\rm T}\rho}{r^2} \left ( 1 + \Pi + \frac{p}{\rho} + \frac{2m_{\rm T}}{r} + 4\pi \frac{r^3p}{m_{\rm T}} \right )
+ {\cal O}(2\mbox{PN}),
\label{dpdr_1PN}
\\
& \frac{dm_{\rm T}}{dr} = 4\pi r^2 \rho ( 1 + \Pi).
\label{dmdr_1PN}
\end{align}
\end{subequations}
where we have introduced the baryonic rest-mass density $\rho$ and the
dimensionless internal energy per unit mass,
$\Pi \equiv (\epsilon -\rho)/\rho$.  It can be noticed that the mass
function equation only contains 1PN corrections to the Newtonian
equations of hydrostatic equilibrium, while higher-order corrections
appear in the pressure equation.


\subsection{The PPN stellar structure equations}
\label{sec:oldPPN}

The PPN formalism \cite{Will:1972zz,Nordtvedt:1972zz} was first
employed for building static, spherically symmetric models of compact
stars by Wagoner \& Malone \cite{WagonerMalone:1974}, and subsequently
by Ciufolini \& Ruffini \cite{CiufoliniRuffini:1983}. This early work
is briefly reviewed here since it will provide the stepping stone
towards formulating our post-TOV equations.

A convenient starting point is the set of stellar structure equations
derived in~\cite{CiufoliniRuffini:1983} from the original
Will-Nordtvedt PPN theory \cite{Will:1972zz,Nordtvedt:1972zz}. These
are [cf. Eqs.~(14) of~\cite{CiufoliniRuffini:1983}]:
\begin{subequations}
\label{PPN_CR}
\bear
&& \frac{dp}{dr} = -\frac{\epsilon \bar{m}}{r^2} \left [1 + (5+3\gamma -6\beta + \zeta_2) \frac{\bar{m}}{r} + \frac{p}{\epsilon}
+ \zeta_3 \frac{E}{\bar{m}}
\label{dpdr1}
 \right.
\nonumber  \\
&& \left. + (\gamma+\zeta_4) \frac{4\pi r^3 p }{\bar{m}} + \frac{1}{2} (11+\gamma-12\beta +\zeta_2 -2\zeta_4 )\frac{\Omega}{\bar{m}}  \right ],
\nonumber \\
\\
&& \frac{d\bar{m}}{dr} = 4\pi r^2 \epsilon,
\label{dmdr1}
\eear
\end{subequations}
where we have adopted the standard notation for the nine PPN
parameters,
$\{\beta,\gamma,\zeta_1,\zeta_2,\zeta_3,\zeta_4,\alpha_1,\alpha_2,\alpha_3\}$. In
the GR limit $\beta=\gamma=1$ and $\zeta_i = \alpha_i =0$
($i=1,...,4$) \cite{Will:2014xja}.

It should be pointed out that the basic parameters $p,\bar{m}$ (as well as the radial coordinate $r$) entering Eqs. (\ref{PPN_CR})
may not be the same as the corresponding ones in the TOV equations. This is a reflection of the ``gauge'' freedom in defining these
parameters in a number of equivalent ways. Indeed, below we are going to exploit  this freedom and obtain
an ``improved'' set of PPN equations by a suitable redefinition of the mass function.  On the other hand, following~\cite{CiufoliniRuffini:1983},
we will stick to the  same $p$ and $r$ throughout this analysis, implicitly assuming that they are the \textit{same} variables as the ones in the TOV equations~(\ref{TOV}).

The potentials $\Omega$ and $E$ appearing in Eq.~(\ref{dpdr1}) obey
\be
\frac{d\Omega}{dr} = -4 \pi r \rho \bar{m}, \qquad \frac{dE}{dr} = 4\pi r^2 \rho \Pi .
\ee
The more familiar Newtonian gravitational potential $U$, solution of
$\nabla^2U=-4\pi\rho$, is not featured in Eqs.~(\ref{PPN_CR}) as a
result of a change of radial coordinate and a redefinition of the mass
function $\bar{m}$ with respect to the original PPN theory parameters
(see~\cite{CiufoliniRuffini:1983} for details).

The stellar structure equations can be manipulated further by switching to a new mass function:
\be
m (r) = \bar{m}  + A E + B \Omega + C \frac{\bar{m}^2}{r} + D (4\pi r^3 p),
\label{meq}
\ee
where $A$, $B$, $C$, and $D$ are free constants. As evident, $\bar{m}$ and $m$
differ at 1PN level. The constants $A$ and $B$ can be chosen so that
the terms proportional to $E$ and $\Omega$ in Eq.~(\ref{dpdr1}) are eliminated.
This is achieved for
\be
A = \zeta_3, \qquad B = \frac{1}{2} ( 11 +\gamma -12\beta + \zeta_2 -2 \zeta_4).
\label{A+B}
\ee

The resulting ``new'' set of PPN stellar structure equations is
\begin{subequations}
\label{newPPN}
\bear
&& \frac{dp}{dr} = -\frac{\rho m}{r^2} \left [ 1 + \Pi +  \frac{p}{\rho} + \left ( 5 + 3\gamma -6\beta + \zeta_2 -C \right ) \frac{m}{r} \right.
\nonumber \\
&& \left. \qquad +~ \left ( \gamma + \zeta_4 -D \right ) 4\pi \frac{r^3 p}{m} \right ],
\label{dpdr_gen}
\\
\nonumber \\
&& \frac{dm}{dr} = 4\pi r^2 \rho \left [ 1+ (1+\zeta_3)\Pi + 3D \frac{p}{\rho} -\frac{C}{4\pi} \frac{m^2}{\rho r^4}
\right. \nonumber \\
&& \left. \qquad  -\frac{1}{2} \left (11+\gamma-12\beta +\zeta_2 -2\zeta_4 -4C +2D \right ) \frac{m}{r}  \right ]. \nonumber \\
\label{dmdr_gen}
\eear
\end{subequations}

These expressions still contain the gauge freedom associated with the definition of the mass function $m$ in the form of the yet unspecified
constants $C$ and $D$. In particular, the Wagoner-Malone hydrostatic equilibrium
equations~\cite{WagonerMalone:1974} represent a special case of these
expressions, and it is straightforward to see that they can be recovered for
\be
D = \gamma + \zeta_4, \qquad C = \frac{1}{2} \left ( 7 + 3\gamma -8\beta + \zeta_2 \right ).
\label{D+C}
\ee
Making this choice for the constants on the right-hand side of
Eq.~(\ref{meq}) leads to a new mass function, say $\tilde{m}$, and to
the following structure equations, which match Eqs.~(6) and (7) of
\cite{WagonerMalone:1974}:
\begin{subequations}
\label{PPN_WM}
\begin{align}
\frac{dp}{dr} &= -\frac{\rho \tilde{m}}{r^2} \left ( 1 + \Pi +  \frac{p}{\rho} + a \frac{\tilde{m}}{r} \right ),
\label{dpdr_WM}
\\
\frac{d\tilde{m}}{dr} &= 4\pi r^2 \rho \left [ 1+ (1+\zeta_3)\Pi +  a \frac{\tilde{m}}{r}  + 3 ( \gamma + \zeta_4)
\frac{p}{\rho}  \right.
\nonumber \\
&\left. \quad  -\frac{b}{4\pi} \frac{\tilde{m}^2}{\rho r^4} \right ],
\label{dmdr_WM}
\end{align}
\end{subequations}
where $a \equiv ( 3 + 3\gamma -4\beta + \zeta_2 )/2$ and the constant
$b$ in the notation of \cite{WagonerMalone:1974} is our $C$, i.e.
$b=\left ( 7 + 3\gamma -8\beta + \zeta_2 \right )/2$.

A comparison between the two sets of PPN equations \eqref{PPN_CR} and
\eqref{PPN_WM} discussed in this section reveals that the
Wagoner-Malone equations are simpler, in the sense that they do not
depend on the auxiliary potentials $\Omega$ and $E$. This advantage,
however, is partially offset by the more complicated expression for
the mass function equation.
If we compare the GR limit of the Wagoner-Malone equations~(\ref{PPN_WM})
against the 1PN expansion of the TOV equations, Eqs~(\ref{TOV1PN}), we
find that the two sets coincide provided we identify
$\bar{m} = m_{\rm T}$, i.e.
\be
\tilde{m} = m_{\rm T} + \frac{m_{\rm T}^2}{r} + 4\pi r^3 p \, ,
\ee
where the last equation follows by taking the GR limit of
Eq.~(\ref{meq}) in combination with Eqs.~(\ref{A+B}) and
(\ref{D+C}). Clearly, the fact that $\tilde{m} \neq m_{\rm T}$ in the
GR limit is an unsatisfactory property of the Wagoner-Malone
equations.

It would be desirable to have a set of structure equations that --
unlike the set \eqref{PPN_CR} -- does not involve integral potentials,
and such that -- unlike the set \eqref{PPN_WM} -- the mass function is
compatible with the GR limit. Fortunately, it is not too difficult to
find a new set of PPN equations for which $m=m_{\rm T}$. In the
following section we will propose an improved set of PPN stellar
structure equations that satisfies these requirements.


\subsection{An improved set of PPN equations}
\label{sec:newPPN}

We can exploit the degree of freedom associated with the constants
$C,D$ in Eqs.~\eqref{newPPN} and produce a new set of PPN equations
that exactly match the 1PN TOV equations in the GR limit with
$m=m_{\rm T}$. It is easy to see that this can be achieved by making
the trivial choice
\be
C=D=0.
\ee
Note that the constants $A$ and $B$ are still given by
Eqs.~(\ref{A+B}).  The resulting PPN equations are
\begin{subequations}
\label{ourPPN}
\begin{align}
\frac{dp}{dr} &= -\frac{\rho m}{r^2} \left [\, 1 + \Pi + \frac{p}{\rho} + \left ( 5 + 3\gamma -6\beta + \zeta_2 \right ) \frac{m}{r} \right.
\nonumber \\
&\left. \qquad +~ (\gamma +\zeta_4) 4\pi \frac{r^3 p}{m} \, \right ],
\label{dpdr_new} \\
\frac{dm}{dr} &= 4\pi r^2 \rho \left [\, 1 + (1+\zeta_3) \Pi   \right.
\nonumber \\
&\left. \qquad - \frac{1}{2} \left ( 11 + \gamma -12\beta + \zeta_2 -2\zeta_4 \right ) \frac{m}{r} \, \right ].
\label{dmdr_new}
\end{align}
\end{subequations}
As advertised, in the GR limit these equations reduce to
Eqs.~(\ref{TOV1PN}) with $m=m_{\rm T}$.  The
same equations will be used in Section~\ref{sec:pT} below in the
construction of the desired post-TOV equations.


\subsection{The physical interpretation of the mass function}
\label{sec:mass}

Within the framework of PPN theory, inertial mass and active/passive
gravitational mass are, in general, distinct notions. In the context of
compact stars, expressions for all three kinds of mass are given in
\cite{CiufoliniRuffini:1983}:
\begin{align}
M_{\rm in} &= \bar{m}(\bar{R}) + \left ( \frac{17}{2} + \frac{3}{2} \gamma -10\beta + \frac{5}{2} \zeta_2 \right ) \Omega (\bar{R}),
\label{Min}
\\
M_{\rm a} &= M_{\rm in} + \left ( 4\beta -\gamma-3 -\frac{1}{2} \alpha_3 -\frac{1}{3} \zeta_1 -2\zeta_2 \right ) \Omega(\bar{R})
\nonumber \\
&+ \zeta_3 E(\bar{R}) - \left (\frac{3}{2} \alpha_3 - 3\zeta_4 + \zeta_1 \right ) P \,,
\\
M_{\rm p} &= M_{\rm in} + \left ( 4\beta -\gamma-3 -\alpha_1 + \frac{2}{3} \alpha_2 -\frac{2}{3} \zeta_1 -\frac{1}{3} \zeta_2 \right )
\nonumber \\
&\times \Omega (\bar{R}) ,
\end{align}
where $\bar{R}$ is the stellar radius associated with the mass
function $\bar{m}(r)$ -- i.e. with the set of equations \eqref{PPN_CR}
-- and
\be
P = 4\pi \int_0^{\bar{R}} dr\, r^2 p.
\ee
is the volume-integrated pressure.

In GR the three masses are of course identical,
$M_{\rm in} = M_{\rm a} = M_{\rm p}$.  As argued in
\cite{CiufoliniRuffini:1983}, any theory conserving the four-momentum
of an isolated system should incorporate the equality of the two
gravitational masses, i.e. $M_{\rm a} = M_{\rm p}$. If adopted, this
equality leads to following three algebraic relations for the PPN
parameters:
\begin{align}
&\zeta_3 = 0,
\\
&\zeta_1 - 3\zeta_4  + \frac{3}{2} \alpha_3 = 0,
\label{eq2}
\\
&\zeta_1 + 3\alpha_1 -2\alpha_2 - 5\zeta_2 -\frac{3}{2} \alpha_3 = 0.
\end{align}
We can subsequently write for the common gravitational mass:
\be
M_{\rm g} = M_{\rm a} = M_{\rm p} = \bar{m}(\bar{R}) + F \Omega(\bar{R}),
\ee
with
\be
F = \frac{1}{2} \left ( 11 +  \gamma -12\beta -\alpha_3 + \zeta_2 -\frac{2}{3} \zeta_1 \right ).
\ee
For our new PPN equations with $C=D=0$ the mass equality $M_{\rm a} = M_{\rm p}$
 implies
\be
m(r) = \bar{m}(r) + \frac{1}{2} \left ( 11 + \gamma -12\beta + \zeta_2 -2\zeta_4 \right ) \Omega(r).
\ee
Then with the help of Eq.~(\ref{eq2}) it is easy to see that
\be
M_{\rm g} = m(\bar{R}) + \left ( \zeta_4 -\frac{1}{2} \alpha_3 -\frac{1}{3} \zeta_1  \right ) \Omega (\bar{R}) = m(\bar{R}).
\ee
If $R$ is the stellar radius associated with our PPN equations
\eqref{ourPPN}, the difference $\delta R = R-\bar{R}$ is a 1PN-order
quantity. We can then approximately write
\be
m(\bar{R}) \approx m(R) - \frac{dm}{dr} (R) \delta R \,.
\ee
However, Eq. (\ref{dmdr_new}) implies that $dm/dr (R)=0$ if
$\rho(R) = 0$ at the stellar surface. This is indeed the case for a
realistic EOS. Therefore, we have shown that at 1PN precision the mass
of the system is given by
\be
M_{\rm g} = m(R).
\ee
This elegant result is one more attractive property of the new PPN
equations.


\section{The Post-TOV formalism}
\label{sec:pT}

The logic underpinning the formalism we are seeking is that of
parametrizing the deviation of the stellar structure equations from
their GR counterparts, thus producing a set of post-TOV equations. As
already pointed out in the introduction, the post-TOV formalism is
merely a useful parametrized framework rather than the product of a
specific, self-consistent modified gravity theory (in the spirit of
PPN theory).  In this sense our formalism is akin to the existing
``quasi-Kerr'' or ``bumpy'' Kerr metrics, designed to study deviations
from the Kerr spacetime in GR~(see
e.g.~\cite{Collins:2004ex,Glampedakis:2005cf,Cardoso:2014rha}).

By design the post-TOV formalism should be a more powerful tool for
building relativistic stars than the PPN framework; after all, the
latter is based on a PN approximation of strong gravity. However (as
will become clear from the analysis of this section), our formalism has
its own limitations, the most important one being the fact that the
deviations from GR are introduced in the form of PN corrections. This
could mean that the structure of compact stars with a high degree of
departure from GR may not be accurately captured by the formalism.


\subsection{Post-TOV equations: 1PN order}
\label{sec:pT1}

The recipe for formulating leading-order post-TOV equations is rather
simple: from a suitable set of PPN hydrostatic equilibrium equations
we isolate the purely non-GR pieces.  These 1PN terms are subsequently
added ``by hand'' to the full general relativistic TOV equations,
hence producing a set of parametrized post-Einsteinian equations. It
should be pointed out that this procedure can only be applied at the
level of 1PN corrections. Higher-order corrections should by sought by
other means, such as dimensional analysis (see Section~\ref{sec:pT2}).

In principle, either set of equations, \eqref{PPN_CR}
\cite{CiufoliniRuffini:1983} or \eqref{PPN_WM}
\cite{WagonerMalone:1974}, could have been used. However, our improved
PPN equations \eqref{ourPPN} seem to be best suited for this task.

Considering Eqs.~\eqref{ourPPN}, we first isolate the terms that
represent a genuine deviation from GR.  These are the terms in the
second line in the following equations:
\begin{subequations}
\bear
&&\frac{dp}{dr} = -\frac{\rho m}{r^2} \left ( 1 + \Pi + \frac{p}{\rho}   + \frac{2m}{r}  +  4\pi \frac{r^3 p}{m} \right )
\nonumber \\
&& \qquad -\frac{\rho m}{r^2} \left [ \left ( 3 + 3\gamma -6\beta + \zeta_2 \right ) \frac{m}{r} + ( \gamma-1+\zeta_4) 4\pi \frac{r^3 p}{m} \right ],
\nonumber
\\
\\
&& \frac{dm}{dr} = 4\pi r^2 \rho \left ( 1 + \Pi \right )
\nonumber \\
&& \quad  +~ 4\pi r^2 \rho \left [\zeta_3 \Pi - \frac{1}{2} \left ( 11 + \gamma -12\beta + \zeta_2 -2\zeta_4 \right ) \frac{m}{r} \right ].
\eear
\end{subequations}
The second step consists of adding the non-GR terms to the TOV
equations \eqref{TOV}.  We obtain (recall that $m=m_{\rm T}$)
\bear
&& \frac{dp}{dr} = -\frac{ (\epsilon + p)}{r^2} \left ( \frac{m  + 4\pi r^3 p}{ 1-2m/r} \right )
-\frac{\rho m}{r^2} \left ( \delta_1 \frac{m}{r} + \delta_2 4\pi \frac{r^3 p}{m} \right ), \nonumber \\ \\
&& \frac{dm}{dr} = 4\pi r^2 \left [ \epsilon +  \rho \left ( \delta_3 \frac{m}{r} + \delta_4 \Pi   \right ) \right ],
\eear
where we have introduced the constant post-TOV parameters:
\bear
&&\delta_1 \equiv  3 (1+ \gamma) -6\beta + \zeta_2, \quad \delta_2 \equiv  \gamma-1+\zeta_4,
\label{delta12}
\\
&& \delta_3 \equiv -\frac{1}{2} \left ( 11 + \gamma -12\beta + \zeta_2 -2\zeta_4 \right ), \quad \delta_4\equiv \zeta_3.
\label{delta34}
\eear
As expected, $\delta_i = 0$ in the limit of GR.

The above equations can be written in a more compact form:
\begin{subequations}
\label{PTOV_1PN}
\begin{align}
\frac{dp}{dr} &= \left (\frac{dp}{dr} \right)_{\rm GR} - \frac{\rho m}{r^2} \left ( \delta_1 \frac{m}{r} + \delta_2 4\pi \frac{r^3 p}{m} \right ),
\label{dpdr_pt1}
\\
\frac{dm}{dr} &=  \left (\frac{dm}{dr} \right)_{\rm GR} +  4\pi r^2 \rho \left ( \delta_3 \frac{m}{r} + \delta_4 \Pi   \right ).
\label{dmdr_pt1}
\end{align}
\end{subequations}

These expressions represent our main result for the
\textit{leading-order} post-TOV stellar structure equations.  They
describe the 1PN-level corrections produced by an arbitrary deviation
from GR that is compatible with PPN theory.  In other words,
Eqs.~(\ref{PTOV_1PN}) encapsulate the stellar
structure physics (at this order) for any member of the PPN family of
gravity theories.

We could in principle introduce other 1PN order terms, in the spirit
of the general parametrized framework of deviating from GR that we
have described in the beginning of this section. But the introduction
of such terms would correspond to either redefinitions of coordinates
and/or the mass function at the 1PN level, as we have already seen, or
deviations from special relativity, which we would prefer not to
include.

Unfortunately, it turns out that Eqs.~(\ref{PTOV_1PN}) are of limited
practical value. As discussed in the executive summary, the modern
limits on the PPN parameters suggest that these corrections are very
close to their GR values, because $\beta, \gamma \approx 1$ and
$\alpha_i, \zeta_i \ll 1$, making all the $\delta_i$ parameters very
small.  We should not therefore expect any notable deviation from GR
at the level of the leading-order post-TOV equations. We verified this
claim by explicit calculations of neutron star stellar models with
different EOSs.

Any significant deviations from compact stars in GR have to be sought
at 2PN order and beyond, where the existing observational limits leave
much room for the practitioner of alternative theories of
gravity. This calls for the formulation of a higher-order set of
post-TOV equations, a task to which we now turn.


\subsection{Post-TOV equations: 2PN order}
\label{sec:pT2}

In this section we shall formulate post-TOV equations with
2PN-accurate correction terms. Unlike the calculation of the preceding
section, we now have to build these equations ``from scratch'', given
that the general PPN theory has not yet been extended to 2PN order.
Inevitably, the procedure for building the various 2PN terms will turn out to be somewhat more complicated
than that of the preceding section, heavily relying on dimensional analysis for constructing these terms
out of the available fluid parameters.
Moreover, at 2PN order we also need to consider terms that involve the
integral potentials $U$, $E$ and $\Omega$ (recall that these were
eliminated at 1PN order by a suitable redefinition of the mass
function). However, as shown numerically and via analytical arguments
in Appendix~\ref{sec:potentials}, the integral potentials can be
approximated to a high precision, and for a variety of EOSs, by simple
linear combinations of the non-integral PN terms. As a result, they do
not have to be considered separately in the post-TOV expansion.

To begin with, we can get an idea of the form of some of the 2PN terms we are looking
for by expanding the TOV equations~(\ref{TOV}) to that order.
Let us first consider the pressure equation (\ref{dpdr_tov}):
\bear
&& \frac{dp}{dr} =  -\frac{\rho m}{r^2} \left [\, \left ( 1 + \Pi + \frac{p}{\rho}   \right ) \left (\, 1 +  \frac{2m}{r}  +  4\pi \frac{r^3 p}{m} \right ) \right.
\nonumber \\
&& \left. \qquad +~  \frac{4m^2}{r^2} + 8\pi r^2 p \, \right ]
+ {\cal O}(\mbox{3PN}).
\label{dpdr_new4}
\eear
As anticipated, all 1PN terms appearing here are also present in our PPN equation~(\ref{dpdr_new}).
The produced 2PN corrections are proportional to the following combinations:
\be
\frac{m^2}{r^2},~ \Pi\frac{m}{r},~ r^2 p,~ \frac{mp}{r\rho},~ \Pi \frac{r^3 p}{m},~ \frac{r^3 p^2}{\rho m}.
\label{2pnterms_1}
\ee
Additional 2PN terms that do not appear in the TOV equations can be
constructed by forming products of the available 1PN terms.
The largest set of 1PN terms can be found in the general PPN
equations~(\ref{newPPN}):
\be
\mbox{1PN}: \quad \Pi,\, \frac{p}{\rho},\, \frac{m}{r}, \, \frac{r^3 p}{m}, \, \frac{m^2}{\rho r^4}.
\label{1pnterms}
\ee
We can observe that all terms, except the last one, also appear
in our final PPN equations~(\ref{ourPPN}).  From these we
can reproduce the set (\ref{2pnterms_1}) as well as the additional 2PN
terms:
\bear
&&\frac{r^6 p^2}{m^2}, ~
\Pi \frac{m^2}{\rho r^4}, ~ \frac{m^3}{\rho r^5}, ~\Pi^2, ~ \Pi \frac{p}{\rho},
\nonumber
\\
&&\underbrace{\frac{p^2}{\rho^2},~\frac{m^4}{\rho^2 r^8},~\frac{m^2 p}{\rho^2 r^4}}.
\label{2pnterms_2}
\eear
We have set apart the last three (underbraced) terms of this set
because, as a result of their $\sim 1/\rho^2$ scaling, these terms
will be discarded. In fact, the same fate will be shared by any term
$\sim \rho^\beta$ with $\beta \leq -2$.

There are various reasons why we believe that this selection rule
should be imposed. In our opinion these reasons are quite convincing,
but they fall short of constituting a watertight argument: in all
fairness, if we had a single truly compelling reason, we would not
need more than one.

The first line of reasoning to exclude the presence of negative powers
of $\rho$ (and of the other fluid parameters) in the PN terms is based
on the regularity of these terms at the stellar surface, where
$p,\rho,\Pi \to 0$ for any realistic EOS. A PN term like the second
one in the underbraced group of the set~(\ref{2pnterms_2}) will lead
to a term diverging as $\sim 1/\rho $ at the stellar surface in the
stellar structure equations, and therefore it is not an acceptable PN
correction. Although this surface regularity argument is powerful, it
obviously works only for terms that do not scale with positive powers
of $p$ or $\Pi$.

The second -- heuristic -- argument applies to gravity theories with the
following (symbolic) structure:
\bear
&& \{ \mbox{geometry}\} = 8\pi T^{\mu\nu},
\\
\nonumber \\
&& \nabla_\nu T^{\mu\nu} = 0 ~ \to ~ \frac{dp}{dr} = (\epsilon+p) \{\mbox{geometry} \},
\label{FE}
\eear
where ``geometry" stands for combinations of the metric and its
derivatives, and the last equation assumes a perfect fluid
stress-energy tensor.  The stress-energy tensor and the right-hand
side of Eq.~\eqref{FE} feature
$\epsilon + p = \rho (1 + \Pi + p/\rho)$ and $p$ {\em linearly}. It
can then be argued that the solution of the field equations for the
metric and its derivatives will display a
\be
\{ \mbox{geometry} \} \sim (\epsilon + \tau p)^n  \sim \rho^n \left ( 1 + \Pi + \tau\frac{p}{\rho} \right )^n
\ee
dependence with respect to the fluid variables (where $\tau$ and $n$ are ${\cal O}(1)$ numbers).
Such a solution should lead to pressure-dependent PN terms of the form:
\be
\mbox{PN term} \sim (r^2 \rho)^{n-1} \left ( \frac{p}{\rho} \right )^k, \qquad k = n, n-1, \dots
\label{PNterm}
\ee
where one $\rho$  factor has been removed and absorbed in the Newtonian prefactor of the structure
equations, while at the same time the $r^2$ factor has been added in order to produce a dimensionless quantity.
A key observation is that the form (\ref{PNterm}) assumes a theory that does \textit{not} depend on \textit{dimensional}
coupling constants. Now, according to (\ref{PNterm}) the highest negative power of $\rho$ corresponds to $k=n$, which
means that the scaling with respect to the density should be:
\be
\mbox{PN term} \sim \rho^\beta, \quad \beta \geq -1
\label{PNterm2}
\ee
Based on these arguments, we deem acceptable those PN terms which
scale with $\rho$ as in~(\ref{PNterm2}). This choice is also
consistent with the previous PPN formulas, see Eqs.~(\ref{newPPN}).  A
similar argument can be used to exclude terms with negative powers of
$p$ and $\Pi$\footnote{A related argument for excluding high powers of
  $1/\rho$ is the following.  By virtue of the field equations, the
  Ricci scalar is usually proportional to the energy density of matter
  (at least in the Newtonian limit, if the modified theory reproduces
  GR in the weak field regime): $R\sim \rho$.  If inverse powers of
  $\rho$ are produced by gravity modifications, they should therefore
  originate from terms $\sim 1/R^n$ in the action of the theory. These
  terms are usually associated with ghosts or
  instabilities~\cite{DeFelice:2010aj}, and therefore their presence
  is problematic.}.

Equation~(\ref{2pnterms_1}) and the top row of Eq.~(\ref{2pnterms_2})
represent a large set of 2PN terms emerging from the expansion of the
TOV equation and from products of the various known 1PN terms.  This
set is large but not necessarily complete. Inevitably, a systematic
approach to the problem of ``guessing'' 2PN terms should involve
dimensional analysis.
To improve readability we relegate our dimensional analysis
considerations to Appendix~\ref{sec:DA}, and here we only quote the
main result.  The \textit{most general} form for 2PN order terms is
given by the dimensionless combination:
\be
\Lambda_2 \sim \Pi^\theta (r^2 p )^\alpha (r^2 \rho)^\beta \left ( \frac{m}{r} \right )^{2- 2\alpha-\beta-\theta},
\label{Lterm}
\ee
where $\alpha$, $\beta$, $\theta$ are integers with
\be
\beta \geq -1, \quad
\ee
while different bounds on $\theta$ and $\alpha$ apply to the two
hydrostatic equilibrium equations:
\bear
&& \frac{dp}{dr}: \quad 0 \leq \theta \leq  2, \quad 0 \leq \alpha \leq  2-\theta,
\label{a_lim2}
\\
&& \frac{dm}{dr}: \quad 0 \leq \theta \leq 3, \quad 0 \leq \alpha \leq 3-\theta.
\label{a_lim3}
\eear

The lower bounds on the three parameters $\alpha$, $\beta$, $\theta$
are dictated by the same considerations discussed below
Eq.~\eqref{2pnterms_2}, namely, regularity at the surface and
consistency with the fact that gravitational field equations of the
general form~(\ref{FE}) are unlikely to generate negative powers
higher than $1/\rho$. The upper bounds on $\alpha$ and $\theta$ are
imposed by the regularity at $r=0$ of the stellar structure terms
arising from $\Lambda_2$ (see Appendix~\ref{sec:DA}).


\begin{figure*}[htb]
\includegraphics[width=0.95\textwidth]{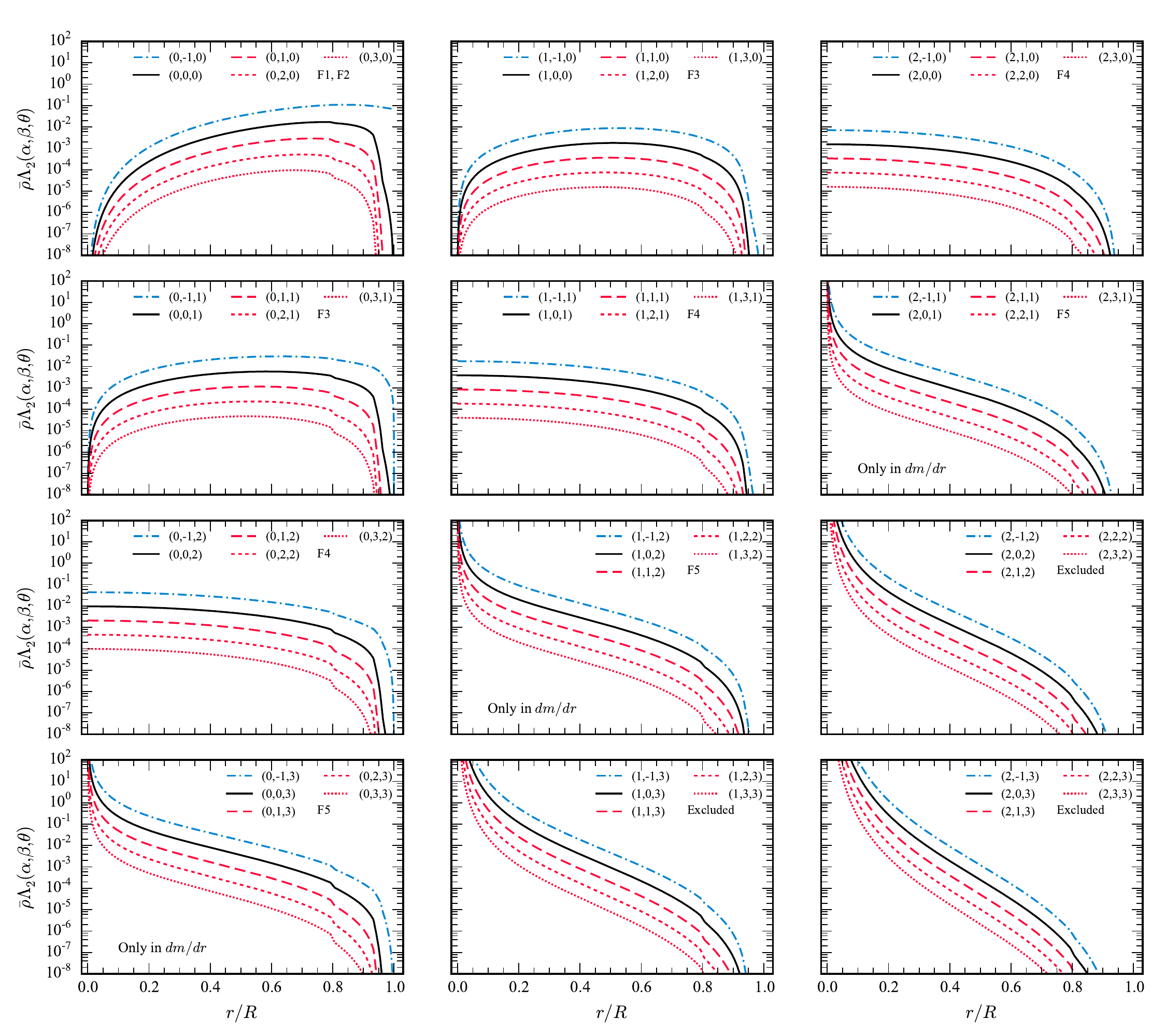}
\caption{ {\it The radial profile of the $\Lambda_2$-function}.  We
  exhibit the behavior of $\bar{\rho} \Lambda_2$, where
  $\bar{\rho} = \rho/\rho_c$ and $\Lambda_2$ is given in
  Eq. (\ref{Lterm}), for a stellar model using the APR EOS, with
  $\epsilon_{\rm c}/ c^2 = 0.86\times 10^{15}$ g/cm$^{3}$,
  $M = 1.51 M_{\odot}$ and $R = 12.3$ km.  The curves are labelled
  according to the respective values of $(\alpha, \beta, \theta)$.
  From the top row to the bottom row the index $\theta$ takes on the
  values $(0, 1, 2, 3)$, respectively. Despite the multitude of
  possible dimensionally correct 2PN terms, their self-similarity --
  which is clear when we compare terms along the bottom-left to
  top-right diagonals in this ``grid'' of plots -- allows us to group
  them into a relatively small number of families (see text for
  details). The contributions plotted in three panels at the bottom
  right of the grid (marked as ``Excluded'') would lead to divergences
  in the hydrostatic equilibrium equations, and therefore they can be
  discarded as unphysical.}
\label{fig:lambdas}
\end{figure*}

From the general expression (\ref{Lterm}) we can reproduce all
previous 1PN and 2PN terms and generate an infinite number of new
ones. This possibility could have been a fatal blow to our post-TOV
program. Fortunately, the day is saved by the fact that the
magnitude of $\Lambda_2$ decays rapidly throughout the star as $\beta$
increases. This trend is clearly visible in the numerical results
shown in Fig.~\ref{fig:lambdas} (see discussion below).
\begin{table}[tbh]
\begin{center}
\begin{tabular}{  p{1.5cm}  p{2.0cm}  p{2.0cm}  p{2.0cm}}
\hline\hline
Family  &  2PN term  & $(\alpha, \beta, \theta)$ &  Dominant /\\
        &            &                           &  Chosen? \\
\hline
F1  &  $m^3/(r^5 \rho)$  & $(0,-1,0)$ &  D/C  \\
\hline
F2  &  $(m/r)^2$  &  $(0,0,0)$ & D/C \\
F2  &  $r m \rho$  &  $(0,1,0)$ & $-$ \\
\hline
F3  &  $mp/(r \rho)$  &  $(1,-1,0)$  & $-$ \\
F3  &  $r^2 p$  &  $(1,0,0)$  & $-$ \\
F3  &  $\Pi m^2/(r^4 \rho)$  &  $(0,-1,1)$  & D/C \\
F3  &  $\Pi m/r$  &  $(0,0,1)$  & $-$ \\
F3  &  $r^2 \Pi \rho$  &  $(0,1,1)$  & $-$ \\
\hline
F4  &  $r^3 p^2 /(\rho m)$  &  $(2,-1,0)$  & $-$ \\
F4  &  $r^6 p^2 /(m^2)$  &  $(2,0,0)$  & $-$ \\
F4  &  $\Pi p/\rho$  &  $(1,-1,1)$  & C \\
F4  &  $\Pi r^3 p/m$  &  $(1,0,1)$  & $-$ \\
F4  &  $\Pi^2 m/(r^3 \rho)$  & $(0,-1,2)$  &  D  \\
F4  &  $\Pi^2$  & $(0,0,2)$  &  $-$  \\
\hline
F5  &  $\Pi r^4 p^2/(\rho m^2)$  &  $(2,-1,1$)  & $-$ \\
F5  &  $\Pi r^7 p^2/m^3$  &  $(2,0,1$)  & $-$ \\
F5  &  $\Pi^2 r p/m \rho$  &  $(1,-1,2)$  &  $-$  \\
F5  &  $\Pi^2 r^4 p/m^2$  &  $(1,0,2)$  &  $-$  \\
F5  &  $\Pi^3/(r^2 \rho)$  &  $(0,-1,3)$  &  D  \\
F5  &  $\Pi^3 r/m$  &  $(0,0,3)$  &  C  \\
\hline\hline
\end{tabular}
\end{center}
\caption{{\it Taxonomy of the dominant 2PN terms.}
  The self-similarity between the radial profiles of the various 2PN terms listed in Eq.~(\ref{2pnterms_3})
  (and  illustrated in Fig.~\ref{fig:lambdas}) allows us to group them into five
  distinct families. This table spells
  out the explicit form of the various terms, and
  indicates which term in each family is dominant (D) according to our
  numerical calculations, and which one was chosen (C) as
  representative of each family.}
\label{tab:classification}
\end{table}

For all practical purposes these results imply that the first few
members of the $\beta = -1, 0, 1, ...$ sequence are sufficient to
construct accurate post-TOV expansions.  A sample set of such dominant
2PN terms is:
\bear
\mbox{2PN}: \quad && \frac{m^3}{r^5 \rho}, ~\frac{m^2}{r^2}, ~ r \rho m, ~ \frac{mp}{r \rho}, ~ r^2 p,
~ \frac{r^3 p^2}{\rho m}, ~\frac{r^6 p^2}{m^2},
\nonumber \\
&& \frac{r^7 p^3}{\rho m^3}\, ~ \frac{r^{10} p^3}{m^4}, ~\Pi \frac{m^2}{r^4 \rho}, ~ \Pi \frac{m}{r},
~ \Pi r^2 \rho, ~\Pi \frac{p}{\rho},
\nonumber \\
&&   \Pi \frac{r^3 p}{m}, \Pi \frac{r^4 p^2}{\rho m^2},  \Pi \frac{r^7 p^2}{m^3},~ \Pi^2 \frac{m}{\rho r^3},
~ \Pi^2,~ \Pi^2 \frac{rp}{m\rho},
\nonumber \\
&&  \Pi^2 \frac{r^4 p}{m^2},~\frac{\Pi^3}{r^2 \rho},~ \Pi^3 \frac{r}{m}.
\label{2pnterms_3}
\eear

This set is markedly larger than the previous sets (\ref{2pnterms_1})
and (\ref{2pnterms_2}) (whose acceptable terms form a subset of the
new set), but a complete post-TOV formalism would have to include all
(or almost all) of these terms, with twice the number of free
coefficient in the $dp/dr$ and $dm/dr$ equations.  Fortunately, as it
turns out, the same job can be done with a much smaller subset of 2PN
terms.  This is possible because the various 2PN terms can be divided
into \textit{five ``families"}, each family comprising terms with
similar profiles. When incorporated in the post-TOV equations, terms
belonging to a given family lead to \textit{self-similar}
modifications in the mass-radius curves for a given EOS.


\begin{figure*}[hbt]
\includegraphics[width=\textwidth]{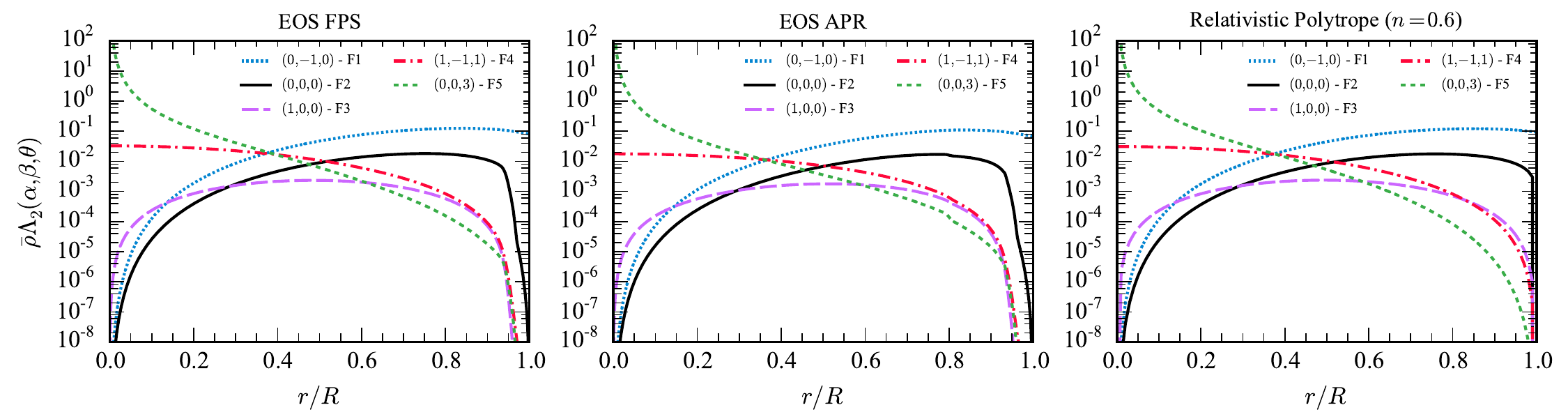}
\caption{{\it The family-representative 2PN terms.}  Here we show the selected representative
  terms from each of the families depicted in Fig.~\ref{fig:lambdas}, as listed in Eq.~(\ref{2pnterms_4}),
  for three different EOSs: FPS (left panel), APR (middle panel)
  and an $n=0.6$ polytrope (right panel).  Each term illustrates the
  qualitative behavior of each family of possible 2PN contributions
  to the structure equations.  The high degree of invariance of the
  $\Lambda_2$-profiles with respect to the EOS is evident in this
  figure.
  The GR background stellar models utilized in the figure have the following bulk
  properties: $\epsilon_{\rm c} = 0.861 \times 10^{15}$ g/cm$^3$ ($\lambda \equiv
  p_{\rm c}/\epsilon_{\rm c} = 0.165$), $M = 1.51 \, M_{\odot}$ and $R = 12.3$ km
  (left panel); $\epsilon_{\rm c} = 1.450 \times 10^{15}$ g/cm$^3$
  ($\lambda = 0.198$), $M = 1.50 \, M_{\odot}$ and
  $R = 10.7$ km
  (center panel);
  $\lambda = 0.165$, $M = 1.50 \, M_{\odot}$ and
  $R = 11.75 $ km
  (right panel).
 }
\label{fig:lambda_dominant}
\end{figure*}


\begin{figure*}[hbt]
\includegraphics[width=\textwidth]{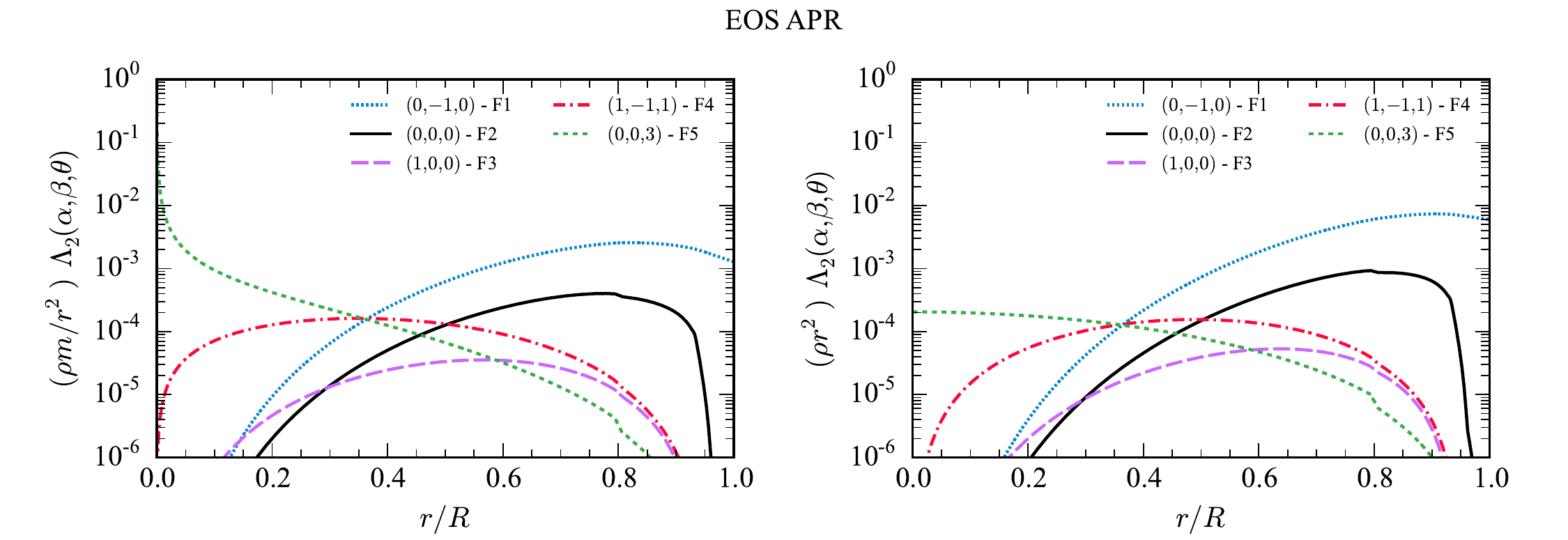} \\
\includegraphics[width=\textwidth]{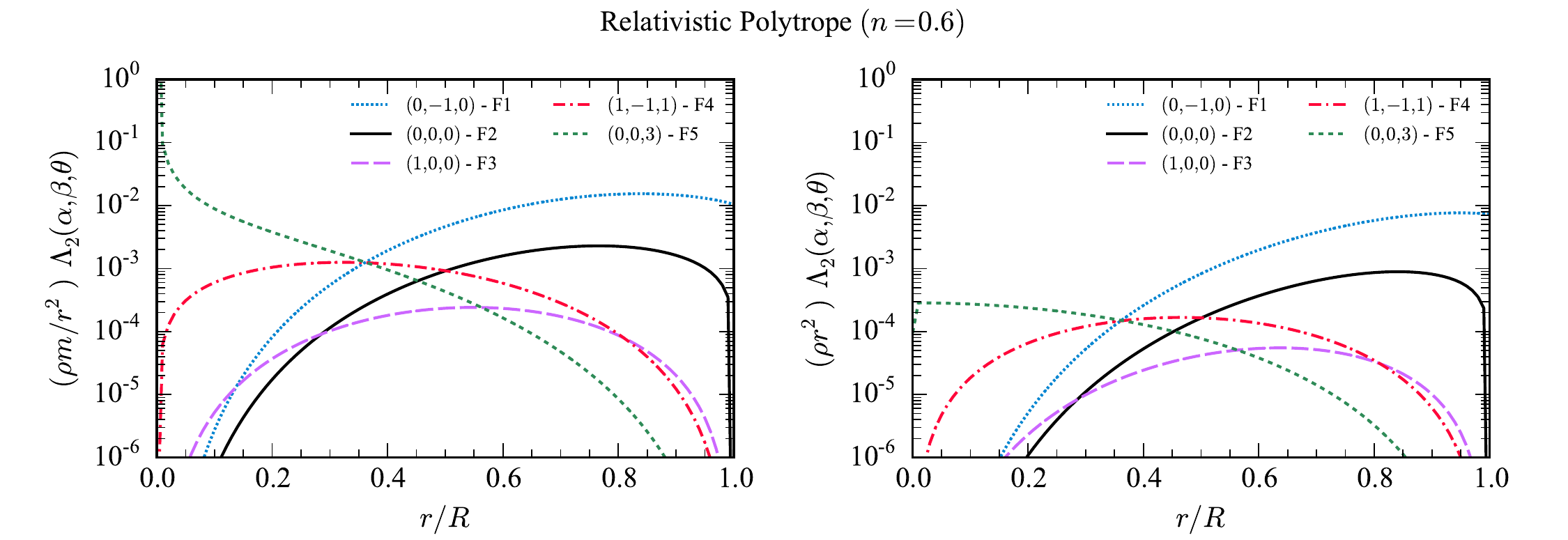}
\caption{{\it The family-representative terms in the structure equations.} This figure
illustrates the behavior of each of the family-representative 2PN terms [see Eq.~(\ref{2pnterms_4})] multiplied
by the Newtonian prefactors in the post-TOV equations. The stellar parameters are identical to the ones used
in Fig.~\ref{fig:lambda_dominant}.
{\it Left panel:} the combination
$({\rho} {m}/{r}^2)\, \Lambda_2(\alpha, \beta, \theta)$ appearing in
the pressure equation.  {\it Right panel:} the combination
${\rho} {r}^2\, \Lambda_2(\alpha, \beta, \theta)$ appearing in the
mass equation. The top panels correspond to EOS APR; the bottom
panels correspond to a relativistic polytrope with polytropic index
$n = 0.6$.  The divergence at the origin of the F5 term justifies its
exclusion from the pressure equation.  }
\label{fig:lambda_dominant_full_factors}
\end{figure*}

Insight into the behavior of the $\Lambda_2 (\alpha,\beta,\theta)$
terms can be gained by direct numerical calculations of their radial
profiles in relativistic stars. We carried out such calculations for a
variety of realistic EOSs as well as relativistic polytropes, and for different
choices of central density, verifying that all cases lead to very similar results,
as discussed below.
More specifically, we considered EOS
A~\cite{1971NuPhA.178..123P}, FPS~\cite{Friedman:1981qw},
SLy4~\cite{Haensel:1993zw,Douchin:2001sv} and
N~\cite{1979PhLB...86..146S} in increasing order of stiffness, as well
as relativistic polytropes with indices $n = 0.4$, 0.6, and 1.0: see
Appendix~\ref{sec:RLE}, and in particular Eq.~(\ref{eq:poly}).
Note that the polytropic models are parametrized by
$\lambda = p_{\rm c}/\epsilon_{\rm c}$ instead of $\epsilon_{\rm c}$ alone (the subscript ``c" 
indicates a quantity evaluated at the center), but this is equivalent to the central density parametrization.
The polytropic models are also invariant with respect to the scale
factor $K^{n/2}$; this can be adjusted to generate polytropic
models of (say) the same mass (for a given $\lambda$) as that of a specific tabulated-EOS model.

Rather than computing $\Lambda_2$ itself, from a phenomenological
point of view it makes more sense to consider the combination
$\rho \Lambda_2$. The reason is that this combination appears in both
the pressure and mass equations, and furthermore it has the desirable
feature of being regular at the surface for $\beta=-1$.  More
specifically, in Fig.~\ref{fig:lambdas} we plot the dimensionless
combination $\bar{\rho} \Lambda_2(\alpha, \beta, \theta)$, where
$\bar{\rho} = \rho/\rho_{\rm c}$.
Our sample neutron star model was built using the APR EOS with central energy density
$\epsilon_{\rm c}/ c^2 = 0.86\times 10^{15}$ g/cm$^{3}$, corresponding
to mass $M = 1.51 M_{\odot}$ and radius $R = 12.3$ km, but we have
verified that our qualitative conclusions remain the same for
different models and different EOSs.

Figure~\ref{fig:lambdas} reveals two key trends: (i) the clear
$\beta$-ordering of the $\rho \Lambda_2$ profiles, with $\beta=-1$
always associated with the dominant term for fixed $\alpha$ and
$\theta$, and (ii) the remarkable similarity in the shape of the
profiles of terms with dissimilar $(\alpha, \beta, \theta)$ triads
along the bottom-left to top-right diagonals in the ``grid" of
Fig.~\ref{fig:lambdas}. This property defines distinct families of 2PN
terms and implies that the terms of each family cause self-similar
changes in the mass-radius curves of the various post-TOV stellar
models.

We have identified five 2PN families (labeled ``F1'',...,``F5'' in the
various panels of Fig.~\ref{fig:lambdas}, and described in more detail
in Table~\ref{tab:classification}):

(i) F1: This is a single-member family comprising only the
$\rho \Lambda_2 (0, -1, 0)$ term in the top-left panel, which is zero
at $r = 0$ but finite at $r = R$.

(ii) F2: The members of this family vanish at $r = 0$ and $r = R$, and
have a peak near the surface. These are the
$\rho \Lambda_2 (0,\beta,0)$ terms with $\beta \geq 0$ in the top-left
panel.

(iii) F3: These terms also vanish at both $r = 0$ and $r = R$, but
display an approximately flat profile inside the star. They correspond
to $\rho \Lambda_2 (1, \beta, 0)$ (top-middle panel) and
$\rho \Lambda_2 (0, \beta, 1)$ (top-right panel) for $\beta \geq -1$.

(iv) F4: This family comprises terms that are finite at $r = 0$ but
zero at $r = R$. These are the $\rho \Lambda_2 (2,\beta,0)$
(bottom-left panel) and $\rho \Lambda_2 (1,\beta,1)$ (bottom-middle
panel) terms with $\beta \geq -1$.

(v) F5: These terms by themselves diverge at $r = 0$ and vanish at
$r = R$, but they become well-behaved when inserted into the stellar
mass-function equation, where they are multiplied by the factor $r^2$:
cf. Eq.~\eqref{dmdr_r0}.
These terms correspond to $\rho \Lambda_2 (2, \beta, 1)$, and from the 
constraints (\ref{a_lim2}) and (\ref{a_lim3}) we conclude that {\em the members of
  this family can only appear in the mass equation}.

There is an intuitive way to explain the existence of the above
families.  As an example we consider F3, where the seemingly
unrelated terms $\Lambda_2 (1,\beta,0)$ and $\Lambda_2 (0,\beta,1)$
yield similar profiles. Consider
\be
\Lambda_{2}(0,\beta,1) \sim r^{-1 + 3\beta} \frac{\Pi \rho^{\beta}}{m^{\beta - 1}}.
\ee
By means of the approximations $m \sim \rho r^3,~\Pi \sim p/\rho$ (the
latter approximation is motivated by the exact thermodynamical
relation $\Pi = n p/\rho$ for relativistic polytropes with index $n$,
see Appendix~\ref{sec:RLE}) we find
\be \label{Fequiv}
\Lambda_{2}(0,\beta,1) \sim r^{2 + 3\beta} p \left(\frac{\rho}{m}\right)^{\beta}
\sim \Lambda_{2}(1, \beta, 0).
\ee
Similarly we can show that
$\Lambda_{2}(2,\beta,0) \sim \Lambda_{2}(1,\beta,1)$ for the F4 family.
The argument can be generalized to show that terms along the diagonals
of Fig.~\ref{fig:lambdas} are equivalent.

Table~\ref{tab:classification} summarizes the taxonomy of the most
important terms of each family according to the above criteria. The
impact of each of these terms as a post-TOV correction has been tested
for a variety of EOSs.  The results reveal that the members of a given
family lead to self-similar modifications to the stellar mass-radius
curves.  A sample of these numerical results is shown in
Fig.~\ref{fig:MR_families}, which is further discussed in
Section~\ref{sec:num} below.

This remarkable self-similarity property means that we can simply
select one term from each family and emulate the effect of
\textit{all} significant 2PN terms of the same family by simply varying the post-TOV coefficient
associated with the selected term.

In doing so, it is reasonable to choose the simplest terms as family
representatives.  For families F2 and F4 the simplest terms also
happen to be the dominant ones (i.e., the ones with the largest
$\rho \Lambda_2$), while for F3 and F5 they are the first subdominant
ones.  The case of the single-member family F1 is trivial. The five
terms we select based on this reasoning are:
\be
\mbox{F-representatives}:
~\frac{m^3}{r^5 \rho},~ \frac{m^2}{r^2 },~ r^2 p, ~ \Pi \frac{p}{\rho},~
\Pi^3 \frac{r}{m}.
\label{2pnterms_4}
\ee
The phenomenologically relevant radial profiles of
$\bar{\rho} \Lambda_2 (\alpha,\beta,\theta)$ produced by these terms
are shown in Fig.~\ref{fig:lambda_dominant} for three choices of EOS:
FPS, APR and an $n=0.6$ polytrope. The most striking feature of this
figure is the close resemblance of the $\Lambda_2$ profiles of
identical $(\alpha, \beta, \theta)$ triads for different EOSs, which
lends support to the EOS-independence of our selection of post-TOV
terms.

The family-representative terms (\ref{2pnterms_4}) are again shown in
Fig.~\ref{fig:lambda_dominant_full_factors}, where we plot the
combinations that appear in the $dp/dr$ and $dm/dr$ equations, i.e.
$ m \rho \Lambda_2 /r^2$ and $r^2 \rho\Lambda_2$, respectively (in the
latter term we have omitted a trivial prefactor of $4\pi$).  We
consider two different EOSs: APR and an $n=0.6$ polytrope.
All terms displayed are regular at both $r=0$ and $r=R$ with the
exception of the F5 term in the $dp/dr$ equation, which is divergent
at $r=0$ and must be excluded. Once again, the variations in the
radial profiles due to considering different EOSs are extremely mild.

We have thus obtained a \textit{minimum} set of representative 2PN
terms, listed in Eq.~(\ref{2pnterms_4}), which in reality encompasses
a much larger set, like the one obtained from the combination of
Eqs.~(\ref{2pnterms_2}) and (\ref{2pnterms_3}), as well as terms that
involve the integral potentials.

After this admittedly tedious procedure we can finally assemble our
2PN-order post-TOV equations for the pressure and the mass. These are
(omitting the negligibly small 1PN corrections):
\begin{subequations}
\label{PTOV_2PN}
\begin{align}
\frac{dp}{dr} &= \left (\frac{dp}{dr} \right )_{\rm  GR}  -\frac{\rho m}{r^2} \left (\, \pi_1 \frac{m^3}{r^5\rho}
+ \pi_2 \frac{m^2}{r^2} \right.
 \nonumber \\
 &\left. + ~ \pi_3 r^2 p+ \pi_4 \Pi \frac{p}{\rho}
 \, \right ),
\label{dpdr_pt2}
\\
\nonumber \\
\frac{dm}{dr} & = \left ( \frac{dm}{dr} \right )_{\rm GR} + 4\pi r^2 \rho \left (  \mu_1 \frac{m^3}{r^5\rho}
+ \mu_2 \frac{m^2}{r^2}
\right.
\nonumber \\
&\left.  +~ \mu_3 r^2 p + \mu_4 \Pi \frac{p}{\rho} + \mu_5 \Pi^3 \frac{r}{m} \right ).
\label{dmdr_pt2}
\end{align}
\end{subequations}
where, as anticipated in the executive summary, $\pi_i$
($i=1,\dots,4$) and $\mu_i$ $(i=1,\dots,5)$ are free parameters
controlling the size of the corresponding departure from GR.


\subsection{Completing the formalism: the post-TOV metric and stress-energy tensor}
\label{sec:metric}

So far, our post-TOV formalism comprises no more than a pair of
stellar structure equations, Eqs.~(\ref{dpdr_pt2}) and
(\ref{dmdr_pt2}), which can be used for the description of static and
spherically symmetric compact stars.  In this section we show that
there is more to the formalism than meets the eye: to a high
precision it is a ``complete'' toolkit, in the sense that (i) it can
be reformulated in terms of a spherically symmetric metric $g_{\mu\nu}$
and a perfect fluid stress-energy tensor $T^{\mu\nu}$, and (ii) these
two structures are related through the covariant conservation law
$\nabla_\nu T^{\mu\nu}=0$ (where $\nabla_\nu$ is the metric-compatible
covariant derivative), hence respecting the equivalence
principle. Remarkably, it also turns out that the metric and matter
degrees of freedom can be related as in GR, which implies that the
post-TOV formalism is \textit{equivalent} to stellar structure in GR
with a \textit{gravity-modified} EOS for matter and an
\textit{effective} spacetime geometry.

In order to establish the above statements we begin with the following
general result. Assume the static spherically symmetric metric
\begin{align}
ds^2 &= g_{\mu\nu} dx^\mu dx^\nu\nonumber\\
&=  -e^{\nu(r)} dt^2 + \left (1-\frac{2 {\cal M}(r)}{r} \right )^{-1} dr^2 + r^2 d\Omega^2
\label{metric}
\end{align}
and a perfect-fluid stress-energy tensor (with energy density $\cE$ and pressure $\cP$)
\be
T^{\mu\nu} = ({\cal E} + {\cal P}) u^\mu u^\nu + {\cal P} g^{\mu\nu}.
\label{T1}
\ee
For a static spherical fluid ball, the energy-momentum conservation equation
\be
\nabla_\nu T^{\mu\nu} = 0
\label{divT1}
\ee
leads to
\be
\frac{d{\cal P}}{dr} = -({\cal E} + {\cal P}) \Gamma^t_{rt} = -\frac{1}{2} (  \cE + {\cal P} ) \frac{d\nu}{dr}.
\label{dPeff}
\ee
As long as we consider theories respecting (\ref{divT1}) with a
metric-compatible covariant derivative, this result is independent of
the gravitational field equations.

For the mass function $\cM(r)$ we can always write a relation of the form
\be
\frac{d{\cal M}}{dr} = 4\pi r^2 \cE \left[1+Z(r)\right],
\label{dMeff}
\ee
where $Z(r)$ is a theory-dependent function.
Einstein's theory is recovered by setting $Z=0$, as required by the
field equations of GR.

To establish the properties described at the beginning of this section
we will show that we can successfully map our post-TOV equations onto
Eqs.~(\ref{dPeff}) and (\ref{dMeff}) (with $Z=0$).

The full post-TOV equations, Eqs.~\eqref{PTOV_2PN}, can be written in the form:
\bear
&& \frac{dp}{dr} = -\frac{(\epsilon + p)}{r^2} \Gamma(r) -\frac{\rho
  m}{r^2} \left [ \, \left ( 1 + \Pi + \frac{p}{\rho}  \right ) \cP_1 + \tilde{\cal P}_2  \, \right ],
\nonumber\\
\\
&& \frac{dm}{dr} = 4\pi r^2 \epsilon + 4\pi r^2 \rho \left [ \, {\cal M}_1 + {\cal M}_2 \, \right ],
\eear
where $\cP_1, \cP_2$ have been defined in Eqs.~(\ref{PandM}),
\be
\tilde{\cP}_2 \equiv \cP_2 -\left (\Pi + \frac{p}{\rho} \right ) \cP_1
\ee
is a 2PN-order term, and $\Gamma(r) \equiv (m+ 4\pi r^3 p)/(1-2m/r)$.

Based on these expressions, we can define the effective energy density
\be
\epsilon_{\rm eff} \equiv \epsilon + \rho ({\cal M}_1+{\cal M}_2),
\label{Eeff}
\ee
which implies
\be
\frac{dm}{dr} = 4\pi r^2 \epsilon_{\rm eff}.
\ee
Using (\ref{Eeff}) in the pressure equation we have
\begin{align}
\frac{dp}{dr} &= - \left [ \epsilon_{\rm eff} + p -\rho ({\cal M}_1+{\cal M}_2) \right ]\frac{\Gamma}{r^2}\nonumber
\\
 & -\frac{m}{r^2} \left [ \, (\epsilon + p) \cP_1 + ( \epsilon_{\rm eff} + p) \tilde{\cal P}_2  \, \right ],
\end{align}
where we have used the fact that in any 2PN term we can replace the factor $\rho$
by $\epsilon_{\rm eff} + p$, at the cost of introducing 3PN terms. Using (\ref{Eeff}) once more
in the last term, and after some rearrangement, we obtain,
\begin{align}
\frac{dp}{dr} &= - \frac{(\epsilon_{\rm eff} + p)}{r^2} \left [
  (1-{\cal M}_2) \Gamma + m ( \cP_1+\tilde{\cal P}_2 -{\cal
  M}_1 \cP_1) \right ]  \nonumber \\
&+ \frac{\rho}{r^2} {\cal M}_1 \Gamma .
\end{align}
Given that ${\cal M}_1 \ll 1$, the last term can be safely omitted and we are left with
\bear
&& \frac{dp}{dr} \approx - \frac{(\epsilon_{\rm eff} + p)}{r^2} \left [ (1-{\cal M}_2) \Gamma
+ m (\cP_1+\tilde{\cal P}_2 -{\cal M}_1 \cP_1) \right ]
\nonumber \\
\\
&& \quad ~\approx - \frac{(\epsilon_{\rm eff} + p)}{r^2} \left [ (1-{\cal M}_2) \Gamma + m \cP_2  \right ],
\eear
which is of the form (\ref{dPeff}). Note that in this and the
following expressions the small $\cM_1, \cP_1$ terms can be omitted.

The resulting mapping is:
\be
\cP = p, \qquad \cM = m, \qquad \cE=  \epsilon_{\rm eff}.
\ee
It follows that the effective post-TOV metric is
\be
g_{\mu\nu} = \mbox{diag} [\, -e^{\nu(r)}, (1-2m(r)/r)^{-1}, r^2, r^2 \sin^2\theta\,],
\label{gpt}
\ee
with
\bear
&& \frac{d\nu}{dr} \approx \frac{2}{r^2} \left [\, (1-{\cal M}_2) \Gamma + m ( \cP_1+\tilde{\cal P}_2 -{\cal M}_1\cP_1) \, \right ]
\\
&& \quad ~ \approx  \frac{2}{r^2} \left [\, (1-{\cal M}_2) \Gamma + m \cP_2 \, \right ].
\label{dnu}
\eear
From this result we can see that $r$ represents the circumferential radius
of the $r=\mbox{constant}$ spheres and therefore the post-TOV radius $R$
(where $p(R)=0$) coincides with the circumferential radius of the star.

Finally, the effective post-TOV stress-energy tensor is
\be
T^{\mu\nu} = (\epsilon_{\rm eff} + p) u^\mu u^\nu + p g^{\mu\nu},
\label{Tpt}
\ee
and it is covariantly conserved with respect to the metric (\ref{gpt}).

These expressions clearly demonstrate that our post-TOV formalism is completely
equivalent to GR with an effective EOS:
\bear
&& p(\epsilon) \to p(\epsilon_{\rm eff}),
\\
&& \epsilon_{\rm eff}  \approx \epsilon + \rho {\cal M}_2.
\label{eos_eff}
\eear
As is evident from this last expression, $\epsilon_{\rm eff}$ represents a gravity-shifted parameter
with respect to the physical energy density $\epsilon$.
This result highlights a key characteristic of compact relativistic stars when studied in
the context of alternative theories of gravity, namely, the intrinsic degeneracy between
the physics of the matter and gravity sectors.

Whether the above effective description (and in particular its
effective geometry part) can give observables that have a
correspondence to observables of an underlying theory or not depends
on the nature of that theory. As long as the underlying theory admits
a PN expansion, the physical description that arises from the
effective formalism should match that of the physical theory. This
non-trivial issue will be further discussed elsewhere \cite{ptov2}.


\begin{figure*}[ht]
\includegraphics[width=\textwidth]{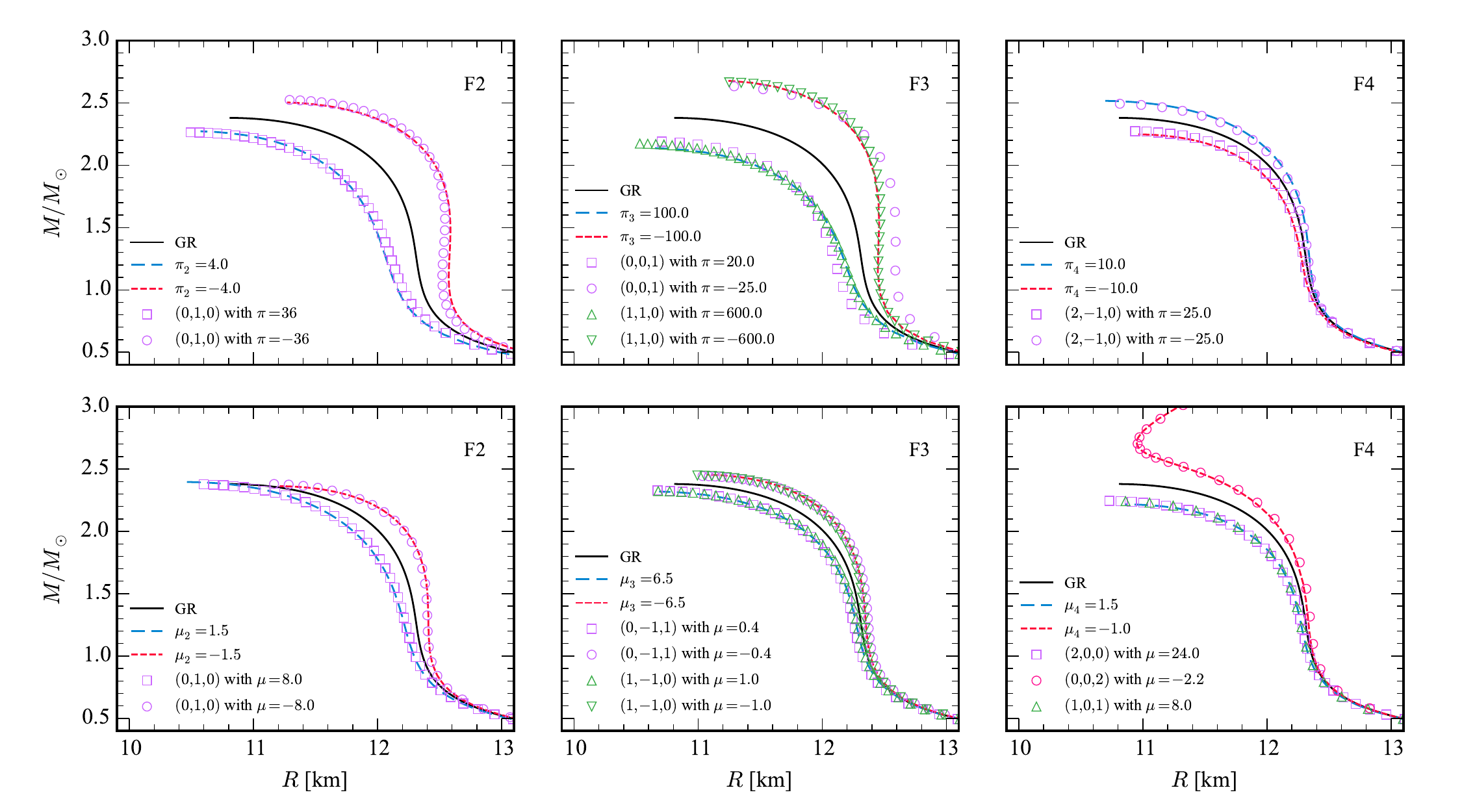}
\caption{{\it Self-similarity in mass-radius curves - I.} Numerical
  integrations show that 2PN terms belonging to the same family result
  in self-similar deviations from GR in the mass-radius relation. This
  figure illustrates this remarkable property for pressure terms (top
  row) and mass terms (bottom row) belonging to families F2, F3 and F4
  (from left to right). In each panel, the solid line corresponds to
  GR; the long-dashed (online: blue) line to a positive-sign
  correction due to the chosen term in each family; the short-dashed
  (online: red) line to a negative-sign correction due to the chosen
  term in each family. The various symbols show that nearly identical
  corrections can be produced using different terms belonging to the
  same family, as long as we appropriately rescale their post-TOV
  coefficients.}
\label{fig:copycat}
\end{figure*}


\begin{figure}[htb]
\includegraphics[width=0.35\textwidth]{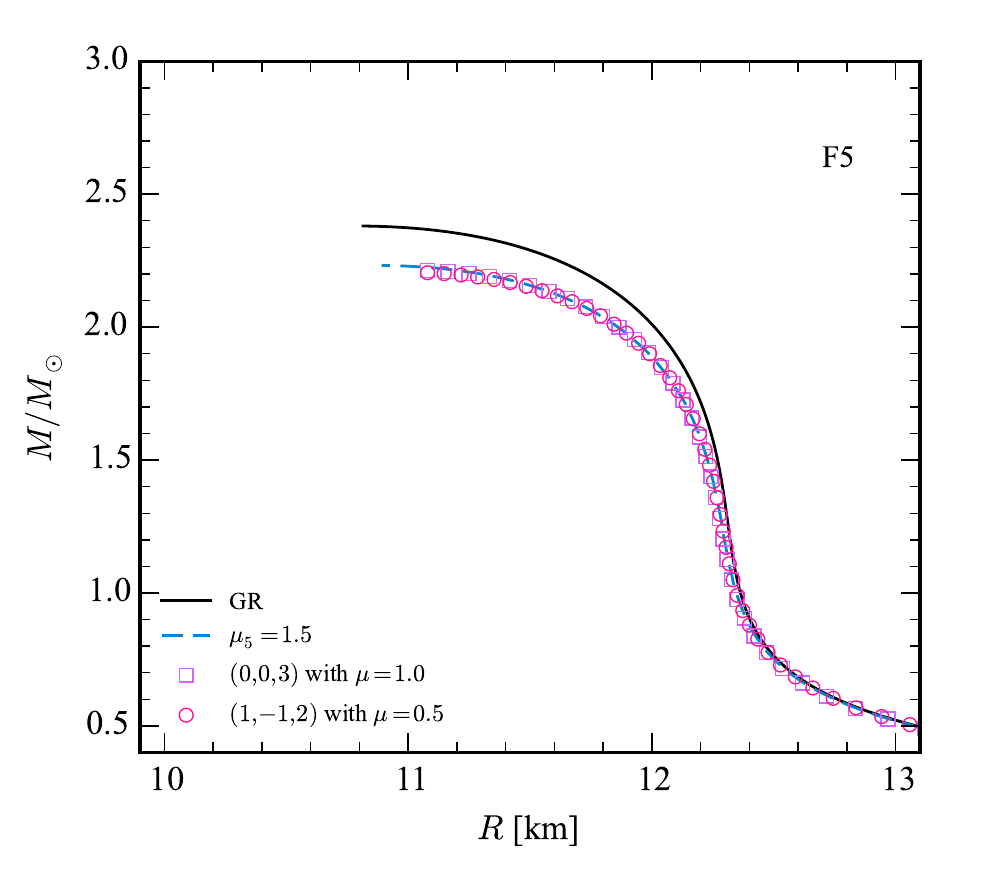}
\caption{{\it Self-similarity in mass-radius curves - II}.  Same as in
  Fig.~\ref{fig:copycat}, but for the F5 family, which only admits
  post-TOV corrections with $\mu_5<0$ (see text).}
\label{fig:mrsm_ex}
\end{figure}


\begin{figure*}[htb]
\includegraphics[width = \textwidth]{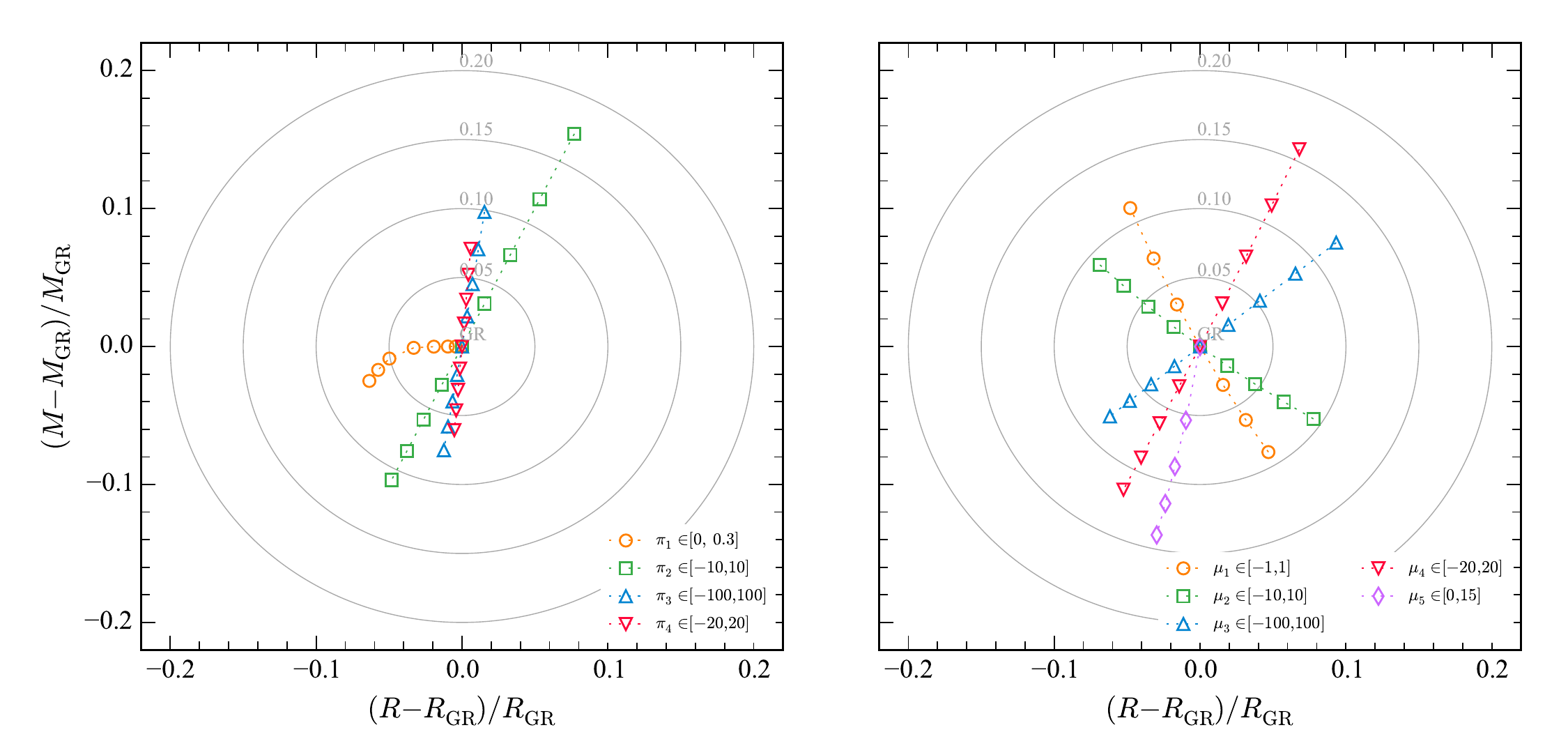}
\includegraphics[width = \textwidth]{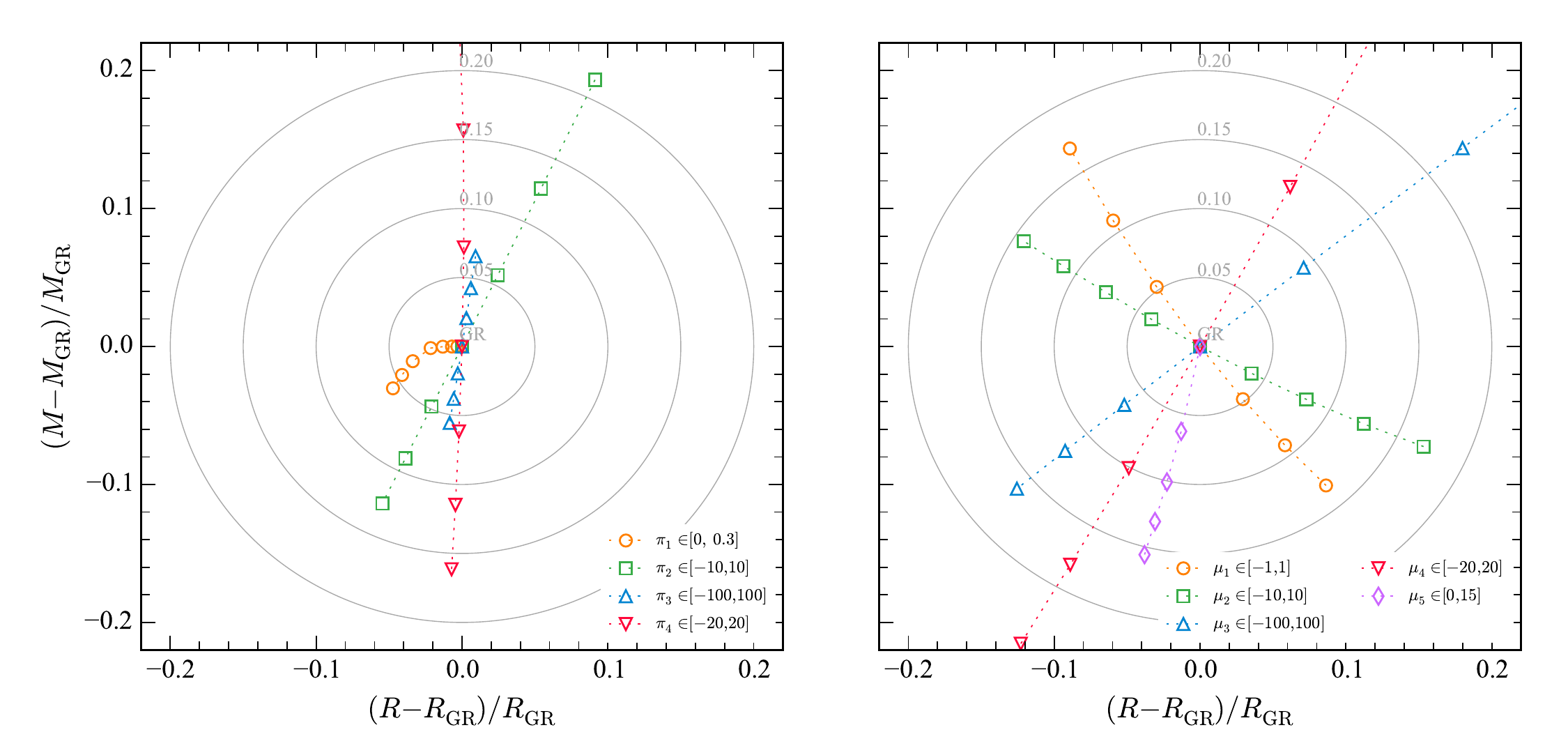}
\caption{{\it Fractional deviations induced by the post-TOV parameters on the stellar mass and radius.}
  Here we illustrate the fractional changes caused by the post-TOV parameters in neutron star masses
  and radii. For a fixed central energy density and EOS APR, we plot
  the relative deviations from GR in mass
  and radius that result from varying the post-TOV parameters within the range indicated in the legends.
  {\it Top row:} $\epsilon_{\rm c}/ c^2 = 8.61\times 10^{14}$ g/cm$^{3}$, $M_{\rm GR} = 1.51~M_{\odot}$
  and $R_{\rm GR} = 12.3$~km.
  {\it Bottom row:} $\epsilon_{\rm c} /c^2 = 1.20 \times 10^{15}$ g/cm$^3$, $M_{\rm GR} = 2.04 \, M_{\odot}$
  and $R_{\rm GR} = 11.9$ km.
  {\it Left panels:} Effect of the post-TOV terms that enter
  in the pressure equation.
  {\it Right panels:} Effect of the post-TOV terms that enter
  in the mass equation.
  The circles represent contours of fixed relative deviation from GR.
}
\label{fig:scatter_apr}
\end{figure*}


\begin{figure}[htb]
\includegraphics[width=0.45\textwidth]{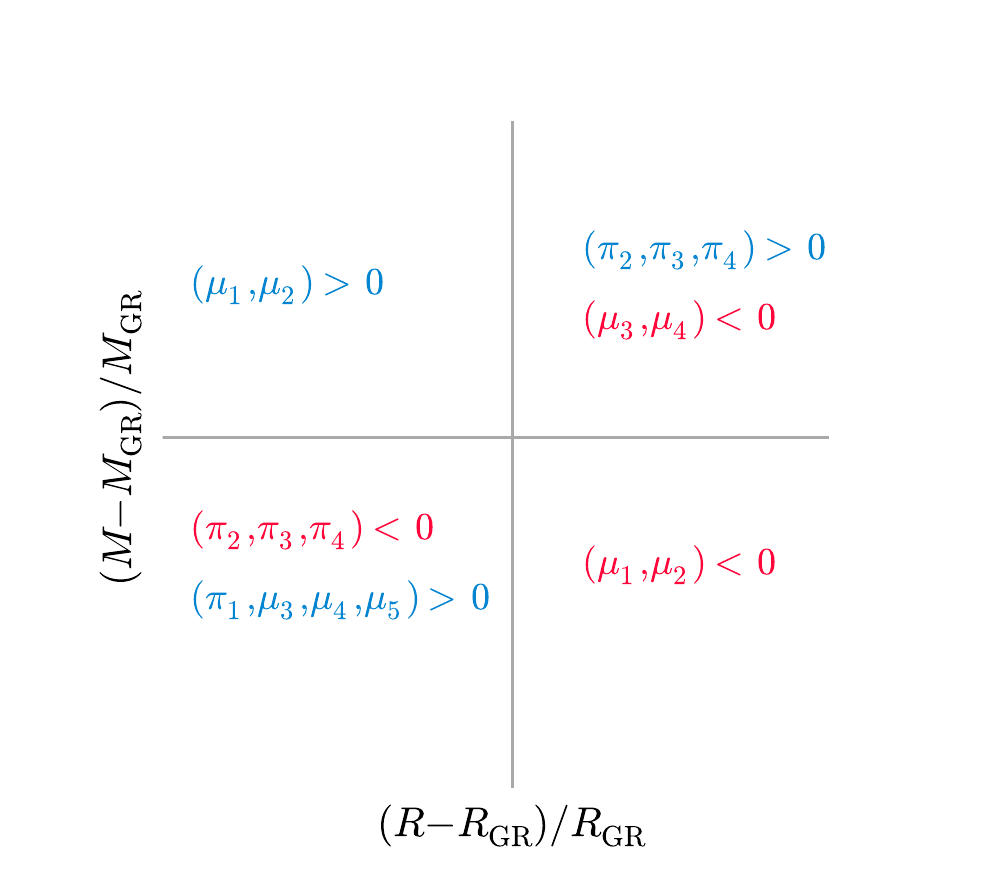}
\caption{{\it Directions of the post-TOV induced deviations.}
This schematic diagram shows which sign of individual post-TOV
parameters produces smaller or larger masses/radii with respect to GR, cf. Fig.~\ref{fig:scatter_apr}.
 }
\label{fig:chart}
\end{figure}


\begin{figure*}[htb]
\includegraphics[width = \textwidth]{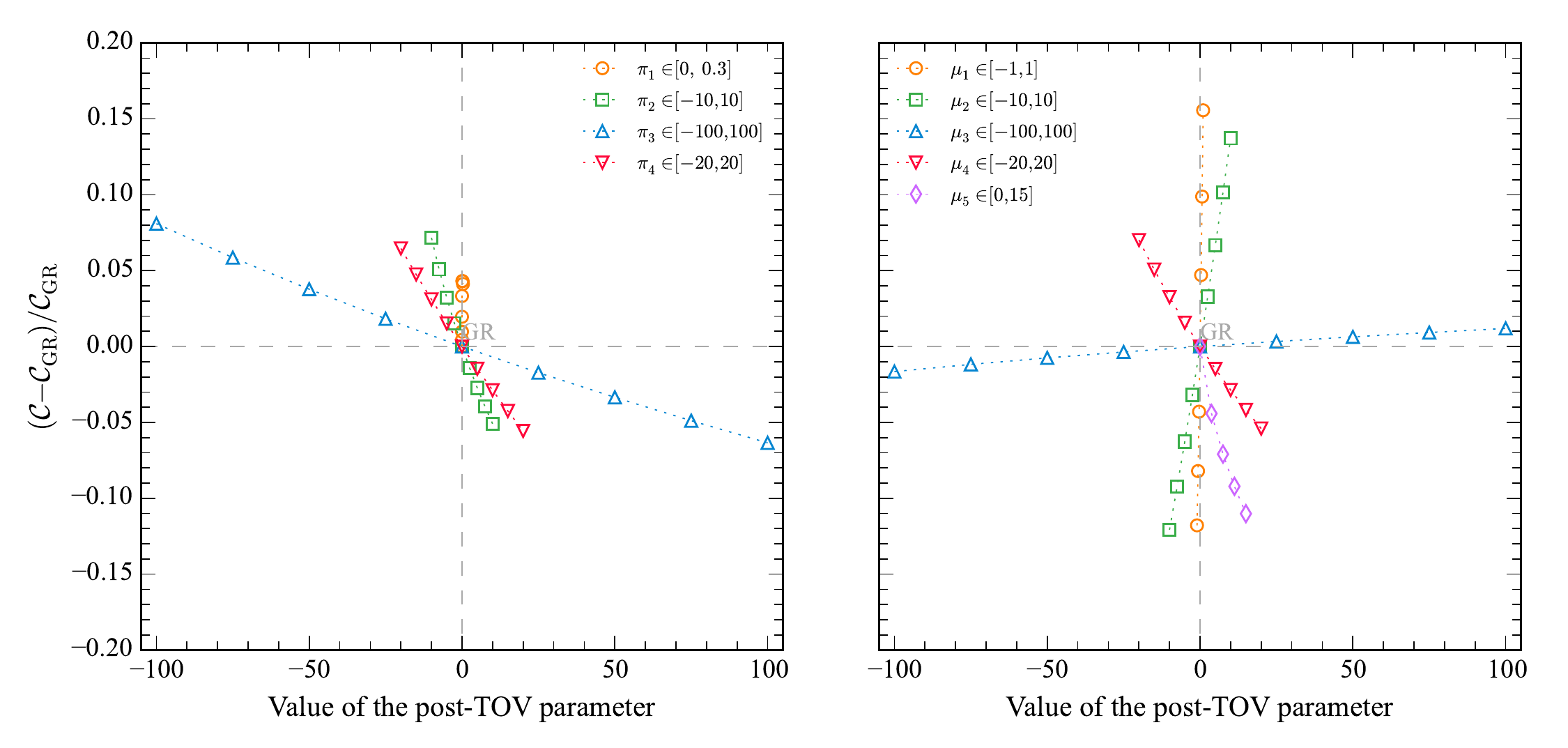}
\includegraphics[width = \textwidth]{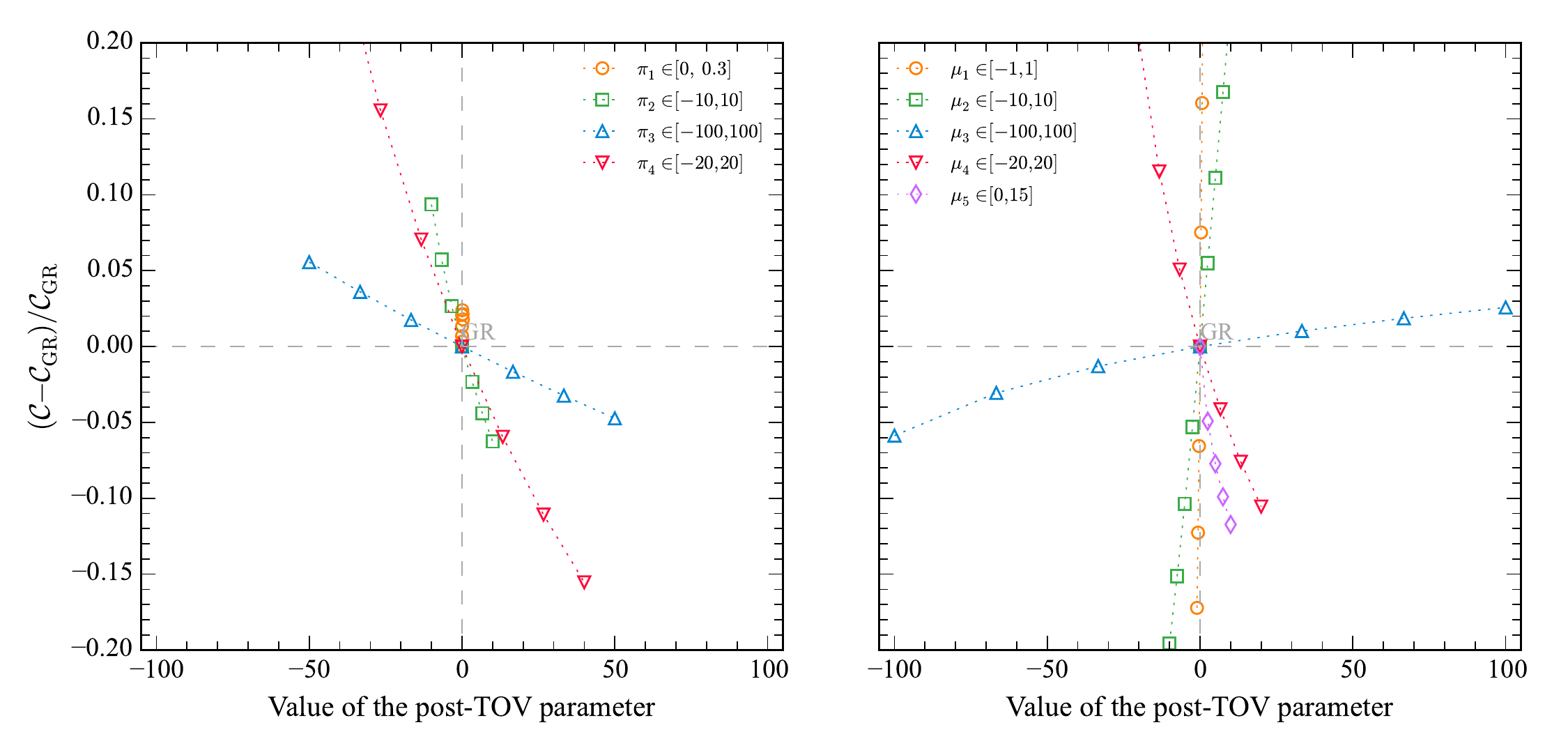}
\caption{{\it Deviations induced by the post-TOV parameters on the
 stellar compactness.}
  Here we consider the influence of the post-TOV parameters on
  the compactness ${\cal C} = M/R$ of neutron stars. Deviations
  from GR are calculated assuming the same APR EOS models
  as in the top and bottom rows of Fig.~\ref{fig:scatter_apr}.
  {\it Left panels:} Effect of the post-TOV terms appearing
  in the pressure equation.
  {\it Right panels:} Effect of the post-TOV terms appearing
  in the mass equation.
  }
\label{fig:scatter_compact_1}
\end{figure*}

\section{Numerical results}
\label{sec:num}

In this section we provide a more detailed discussion of our numerical
techniques and results, focusing on the mass-radius curves produced by
the integration of the post-TOV equations~(\ref{PTOV_2PN}) [or
equivalently Eqs.~(\ref{PTOV_2PN_intro})].

First, let us briefly summarize the integration procedure we have followed in this paper.
We have carried out two kinds of computations: (i) ``background'' models -- these involve the integration of the
general relativistic TOV equations -- with the purpose of studying the radial profiles of the post-TOV correction terms,
(ii) the integration of the full post-TOV equations, typically including the representative term of a single 2PN family.

The post-TOV structure equations~(\ref{eq:PTOV_2PN2_dp}) and (\ref{eq:PTOV_2PN2_dm})
are integrated simultaneously starting at the origin $r = 0$, for fixed values of the coefficients $\pi_i$, $\mu_i$,
and for a range of central energy density values. The chosen central energy density $\epsilon_{\rm c}$
fixes the central pressure $p_{\rm c} = p(\epsilon_{\rm c})$, the central mass density
$\rho_{\rm c} = m_{\rm b} n_{\rm b}(\epsilon_{\rm c})$ and the central
internal energy $\Pi_{\rm c} = (\epsilon_{\rm c} - \rho_{\rm c})/\rho_{\rm c}$, where $m_{\rm b} = 1.66 \times 10^{-24}$~g
is the baryonic mass and $n_{\rm b}$ is the baryon number density. In general, $p(\epsilon)$ and $n_{\rm b}(\epsilon)$
are computed using tabulated EOS data. Once the initial conditions have been specified,
Eqs.~(\ref{eq:PTOV_2PN2_dp}) and (\ref{eq:PTOV_2PN2_dm}) are
integrated outward up to the stellar radius $R$, where $p(R) = 0$.
The gravitational mass is obtained as $M=m(R)$.

The integration procedure for realistic EOS background models is
virtually the same as the one just described.  We have also employed a
number of polytropic background models; for these the integration
procedure is slightly different (see Appendix~\ref{sec:RLE} for
details), and it is based on the simpler Lane-Emden formulation, where
the pressure is replaced by the density $\rho$ in the structure
equations and the stellar model is parameterized by the ratio
$\lambda = p_{\rm c}/\epsilon_{\rm c}$ rather than $\epsilon_{\rm c} $ alone (this
formulation is of course equivalent to the one using tabulated
EOSs). The added advantage of this approach is its scale invariance
with respect to the polytropic constant $K$. This means that $K$ can
be freely adjusted to generate a model with (say) a specific mass
$M$. This scaling procedure also fixes the radius $R$.

The main installment of our mass-radius results has already been
presented in Fig.~\ref{fig:MR_families} of the executive summary
(Section~\ref{sec:summary}).  As discussed there, the various post-TOV
correction terms, representing the five 2PN families of
Section~\ref{sec:pT2}, cause qualitatively different modifications to
the mass-radius curves.

As a rule of thumb, the corrections to the pressure equation lead to
markedly weaker mass-radius modifications than the corrections to the
mass equation, for the same magnitude of $\pi_i$ and $\mu_i$.  The
effective-metric formulation of the post-TOV formalism suggests a
simple qualitative explanation of this observation. The mass
corrections $\cM_2$ change both the effective EOS and the strength of
gravity, as measured by $\nu(r)$, while the pressure corrections
$\cP_2$ are only associated with a change in the strength of gravity
[cf. Eqs.~(\ref{dnu}) and (\ref{eos_eff})], and it is well known that
changes in the EOS outweigh gravity modifications in terms of their
effect on the mass-radius relation.

A notable exception is the single-member family F1, for which the
pressure correction term dominates over its mass counterpart.  In
fact, the F1 pressure term leads to the largest mass-radius changes,
as evidenced by the $\pi_1$ values used in Fig.~\ref{fig:MR_families}.
It is not too difficult to explain why this happens: near the stellar
surface, where all three fluid parameters $p,\rho,\Pi$ are close to
zero, the F1 correction terms remain finite and dominate over all other
terms in the post-TOV equations (this can be clearly seen in
Fig.~\ref{fig:lambda_dominant_full_factors}), thus taking control of
the pressure and mass derivatives.

Another noteworthy point is that, when considering individual post-TOV
terms, it is not always possible to integrate the equations for both
positive and negative values of the corresponding coefficient.  This
is the case for family F5 in Fig.~\ref{fig:MR_families}, where the
integration fails for $\mu_5 <0$.  We have found that this is caused
by an unphysical negative slope $dm/dr$ near the origin.

The remarkable self-similarity in the radial profiles of 2PN terms
belonging to the same family has been illustrated in
Fig.~\ref{fig:lambdas} (see Section~\ref{sec:pT2}). With hindsight,
this property should not come as a total surprise, given the
approximate correlations among the fluid variables: $m\sim \rho r^3$,
$\Pi \sim p/\rho$.

The emergence of the same self-similarity in the mass-radius curves is
something far less anticipated and even more striking.  This property,
which has allowed us to formulate a practical and versatile set of
post-TOV equations, is illustrated in Figs.~\ref{fig:copycat} and
\ref{fig:mrsm_ex}, where we show mass-radius results for each 2PN
family, considering both the pressure and the mass equation and for
the same APR EOS stellar model as in Fig.~\ref{fig:MR_families}.  Each
panel is devoted to a particular family, and it shows the mass-radius
curves resulting from the integration of the post-TOV equations when
various terms from Table~\ref{tab:classification} are included as
corrections (notice that F1 is missing from these plots for the
obvious reason that it consists of only one post-TOV correction).

In all cases considered, the terms of the same family are found to
cause {\em nearly identical mass-radius changes} by a suitable
rescaling of the relevant coefficient $\pi_i$ or $\mu_i$.  This
behavior is most striking for family F4, where different post-TOV
corrections in the mass equations lead to the same characteristic
back-bending behavior in the mass-radius curve.
The only notable exception to this remarkable scaling property is the
$(0,\,0,\,1)$ member of the F3 family, proportional to $\Pi m/r$,
which can be rescaled to agree with other members of the family at
high densities but partially fails to capture the behavior of the
mass-radius curve at low densities. This partial symmetry breaking can
be understood by looking at the leftmost panel in the second row of
Fig.~\ref{fig:lambdas}: the behavior of this term near the surface is
not as smooth as for other members of this family. In our opinion this
does not warrant extensions of the formalism to include another
family, but this is definitely a possibility that could be considered
in the future, given the approximate nature of the self-similarity
argument.

Another important aspect of the post-TOV results is their
``directionality" in the mass-radius plane, in the sense that a given
correction term could affect the mass more than the radius, or vice
versa.  This kind of information cannot be easily extracted from a
traditional mass-radius plot such as Fig.~\ref{fig:MR_families}, but
becomes very visible if we display the same results in terms of the
fractional changes
$ \delta M/M_{\rm GR} \equiv (M-M_{\rm GR})/M_{\rm GR}$ and
$ \delta R/R_{\rm GR} \equiv (R-R_{\rm GR})/R_{\rm GR}$ from the
corresponding GR values.

``Dart-board" plots of these fractional changes are shown in
Fig.~\ref{fig:scatter_apr}.  The aforementioned directionality of the
various post-TOV corrections is clearly visible in this figure.
Individual correction terms are seen to drive nearly \textit{linear}
departures (at least up to a $\sim 10\%$ level) from the center of the
``board." Moreover, certain terms are mutually (nearly) orthogonal,
although not aligned with the mass or radius axis. In some cases this
happens between the pressure and mass terms of the same family,
e.g. family F2. In general, the departures from the GR model are more
isotropically scattered when caused by the corrections $\cM_2$ in the
$dm/dr$ equation, whereas the pressure corrections $\cP_2$ are clearly
more concentrated near the direction of the mass axis. 
This behavior fits nicely with the effective-EOS interpretation of how
$\cM_2$ and $\cP_2$ corrections change the mass-radius diagram. As
expected, $\cM_2$ corrections affect the stiffness of the effective
EOS with significant effects on the radius, while $\cP_2$ corrections
change the strength of gravity, and this mostly affects how much mass
a particular model can support.

These trends remain unchanged as the central energy density (and the stellar mass)
increases (see bottom panels of Fig.~\ref{fig:scatter_apr}).
The pressure correction term associated with $\pi_1$ (family F1) provides the exception to the
rule: a sequence of $\pi_1 >0$ values leads to a non-linear trajectory, with initially just the radius 
decreasing and then followed by a comparable fractional decrease in the mass. 
Negative values of $\pi_1$ are not shown because they lead to unphysical models where in the outer 
low-density layers of the star $dp/dr$ becomes nearly zero but never negative, thus preventing us
from finding the exact location of the surface (as we have pointed out earlier in this 
section, this behavior is related to the non-zero value of the F1 term at the surface).

Fig.~\ref{fig:chart} provides a schematic chart of the correlation
between the sign of the $\pi_i,~\mu_i$ coefficients and the sign of
the associated variations $\delta M$, $\delta R$. Interestingly, the
$\pi_i$ terms are limited to just two of the four possible
quadrants (note the anti-correlation between the signs of $\pi_1$ and the other
$\pi_i$). This translates to variations that simultaneously make the
star bigger (smaller) and heavier (lighter), i.e.  $\delta R>0$,
$\delta M >0$ (or $\delta R<0$, $\delta M <0$). In contrast, the
$\mu_i$ terms occupy all four quadrants, with the F3, F4, F5 families
leading to $\delta R\, \delta M >0$ variations, and F1, F2 giving rise
to the opposite arrangement, $\delta R\, \delta M < 0$.

The linear patterns of Fig.~\ref{fig:scatter_apr} suggest that the
mass and radius variations, for a given $\sigma_i =\{\pi_i \neq \pi_1,\mu_i\}$,
obey the empirical relations,
\be
\frac{\delta M}{M_{\rm GR}}  \approx \sigma_i K_M, \quad \frac{\delta R}{R_{\rm GR}} \approx \sigma_i K_R,
\ee
where the structure parameters $K_M$, $K_R$ are functions of the EOS
and of $\epsilon_{\rm c}$, but they are independent of
$\sigma_i$. Given the nonlinear character of the post-TOV equations,
this conclusion is clearly nontrivial.  We can recast this result in
terms of the variation of the stellar compactness $\cC \equiv M/R$,
\be
\frac{\delta \cC}{\cC_{\rm GR}} \approx \sigma_i  \left ( K_M - K_R \right ).
\ee
This almost linear $\delta \cC (\sigma_i)$ dependence\footnote{It is
  interesting to note that the qualitative effect of the post-TOV
  terms in the pressure equation can be understood by analogy with the
  case of anisotropic stars in GR. The post-TOV pressure equation
  takes the form
  $ dp/dr = \left (dp/dr \right )_{\rm GR} - \rho m \pi_i f_i(r)
  /r^2$,
  with $f_i (r) >0$, whereas anisotropic stars obey
  $dp_r/dr = \left (dp_r/dr \right )_{\rm GR} - 2 \sigma/r$, with
  $\sigma = p_r - p_q$ being the difference between the radial and
  tangential pressure. These two expressions can be matched if
  $2r\sigma = \rho m \pi_i f_i(r)$. The compactness of anisotropic
  stars is known to decrease (increase) when $\sigma$ increases
  (decreases)~\cite{Silva:2014fca}. This conclusion is in good
  qualitative agreement with the results shown in the left panel of
  Fig.~\ref{fig:scatter_compact_1}.}
can indeed be seen in the numerical results shown in
Fig.~\ref{fig:scatter_compact_1}, where we consider the same stellar
models as in Fig.~\ref{fig:scatter_apr}.

The results presented in this section provide a wealth of information on the character of the post-TOV
corrections on stellar structure. It is likely that a more systematic study of the self-similar F-families will
reveal additional layers of information and provide clues as to why the 2PN terms change the bulk properties
of the star the way they do, as a function of the central density. Such a study is beyond the scope of this paper
but provides an attractive subject for future work.


\section{Conclusions and outlook}
\label{sec:concl}

This paper is a first step towards establishing a parametrized
perturbative framework that should, at least in principle, encompass
all modifications to the bulk properties of neutron stars induced by
modified theories of gravity. As in the original formulation of the
PPN formalism, along the way we were forced to make some reasonable
simplifying assumptions in order to reduce the complexity (and
increase the practicality) of our parametrization. These reasonable
assumptions may well fail to match the well-known creativity of
theorists, and it will be interesting to see how the formalism can be
extended and improved.

In a follow-up paper we will use our basic post-TOV equations to
recover stellar structure calculations in some popular theories of
gravity, such as those shown in Fig.~\ref{fig:theory_deg}.  It is
particularly interesting to compare the formalism against theories
that violate some of our basic assumptions, such as scalar-tensor
gravity with spontaneous scalarization (which introduces intrinsically
nonperturbative effects~\cite{Damour:1993hw}) or
Eddington-inspired-Born-Infeld gravity, with its lack of a Newtonian
limit and its unorthodox dependence on the stress-energy
tensor~\cite{Pani:2012qb,Delsate:2012ky}.

We have already obtained some interesting results in this context: for
example, our conclusion that the 2PN post-TOV equations are equivalent
to an effective modified perfect-fluid EOS (see
Section~\ref{sec:metric}) has an interesting parallel with the results
by Delsate et al.~\cite{Delsate:2012ky}, who reached a similar
conclusion for Eddington-inspired-Born-Infeld gravity. We are
currently extending the ``effective metric'' formalism developed in
this paper to the exterior spacetime of compact
stars~\cite{ptov2}. This is necessary to compute physical observables
such as the gravitational redshift of surface atomic lines, the
touchdown luminosity of a radius expansion burst and the apparent
surface area of neutron stars~\cite{Psaltis:2007rv}, and it is
possible that the combination of multiple observables may lift the
EOS/gravity degeneracy.

There are several interesting extensions of our work that should be
addressed in the future. The most obvious one is to assess whether
post-TOV parameters can indeed reproduce the mass-radius curve in
various classes of alternative theories, and whether the post-TOV
parameters encode specific information on the physical parameters
underlying specific theories. This study will hopefully lead to a
better understanding of the generality of the EOS/gravity-theory
degeneracy.

From a data analysis point of view, it is important to understand
whether physical measurements of masses and radii (or perhaps more
realistically, measurements of masses and surface redshifts/stellar
compactnesses) can lead to constraints on the post-TOV parameters
under specific assumptions on the high-density EOS. The answer to this
question obviously depends on the relative magnitude of modified
gravity effects and EOS uncertainties. It will be interesting to
quantify what uncertainties in the EOS are acceptable if we want to
experimentally constrain post-TOV parameters at meaningful levels.

Other obvious extensions are (i) the generalization of the post-TOV
framework to slowly and possibly fast rotating relativistic stars, and
(ii) stability investigations within the post-TOV framework.
We hope that our work will stimulate further activity in this
field. Stability studies in a post-TOV context may reveal that certain
generic features of modified gravity lead to instabilities even for
nonrotating stars, possibly excluding whole classes of modified
gravity theories.

Last but not least, we would like to point out that our post-TOV
toolkit is not (nor was it designed to be) a self-consistent PN expansion, 
but rather a phenomenological parametrization of the
leading-order (unconstrained) deviations from GR. A
systematic and self-consistent PPN expansion extending the PN stellar
structure works cited in the introduction
\cite{Asada:1996ai,Taniguchi:1998wi,Shinkai:1998mg,Gupta:2000zb} is an
interesting but quite distinct area of investigation that should also
be pursued in the future.


\acknowledgments

K.G. is supported by the Ram\'{o}n y Cajal Programme of the Spanish
Ministerio de Ciencia e Innovaci\'{o}n, by the German Science
Foundation (DFG) via SFB/TR7 and by NewCompstar (a COST-funded
Research Networking Programme).
G.P. has received funding from the European Research Council under the
European Union's Seventh Framework Programme (FP7/2007-2013) / ERC
grant agreement n. 306425 ``Challenging General Relativity''.
H.O.S. was supported by NSF CAREER Grant No. PHY-1055103. E.B.  was
supported by NSF CAREER Grant No. PHY-1055103 and by FCT contract
IF/00797/2014/CP1214/CT0012 under the IF2014 Programme.
The authors wish to thank Paolo Pani, Hajime Sotani and Kent Yagi for
discussions and for sharing or validating some of the numerical
results shown in Fig.~\ref{fig:theory_deg}.


\appendix


\section{Dimensional analysis of post-Newtonian terms}
\label{sec:DA}

In this appendix we develop an algorithm for constructing PN terms
using dimensional analysis techniques. Barring PN terms involving the
three potentials $U$, $E$, $\Omega$, the available parameters for
generating PN terms are $\{p,\rho,m,r,\Pi\}$.  From these quantities
plus the gravitational constant $G$ and the speed of light $c$ we can
build the dimensionless combination\footnote{Note that this
  combination is oblivious to the presence of dimensional coupling
  constants that might appear in modified theories of gravity.}:
\be
\Lambda = p^\alpha \rho^\beta m^\gamma r^\delta \Pi^\theta G^\kappa c^\lambda,
\label{da1}
\ee
for a suitable choice of integers
$\alpha,\beta,\gamma,\delta,\kappa,\lambda$ (these are not to be
confused with the PPN parameters of Section~\ref{sec:PPN}).  Since
$\Pi$ is already dimensionless, there is no a priori dimensional
restriction on $\theta$ (apart from one coming from the PN order of
$\Lambda$) and therefore that factor can be omitted in the dimensional
analysis.  Using the scalings
\be
p \sim G \frac{m\rho}{r}, \qquad m \sim \rho r^3,
\ee
we obtain the following form for $\Lambda$ in terms of mass,
length and time dimensions:
\be
\Lambda \sim [M]^{\alpha+\beta+\gamma-\kappa}
[L]^{-\alpha + \delta -3\beta + \lambda + 3\kappa} [T]^{-2\alpha -\lambda
  -2\kappa}.
\label{da2}
\ee
Since $\Lambda$ is required to be dimensionless, we have the three algebraic relations:
\be
\lambda = -2(\alpha +\kappa), \qquad \kappa = \alpha + \beta + \gamma,
\label{da3}
\ee
and
\be
-\alpha + \delta -3\beta + \lambda + 3\kappa = 0.
\label{da4}
\ee
The first two relations simply express $\lambda$ and $\kappa$ in terms of the other parameters. Using them in (\ref{da4})
we obtain
\be
\gamma + \delta = 2 (\alpha + \beta),
\label{da5}
\ee
which represents the true dimensional degree of freedom. It is
straightforward (if tedious) to verify that all PN terms appearing in
the PPN equations of Section~\ref{sec:PPN} are consistent with
(\ref{da5}).

All $\alpha < 0$ terms are divergent at the surface and need not be
considered.  As we shall shortly see, all terms with $\alpha \geq 4$
are divergent at $r=0$ in both structure equations, and therefore
should be discarded. The $\alpha=3$ terms are singular in the $dp/dr$
equation and can be discarded by the same argument; $\alpha=3$ terms
are regular in the $dm/dr$ equation, but they are always dominated in
magnitude by the $\alpha < 3$ terms, and therefore will not be presented in detail here.
Therefore our strategy hereafter is to focus on the
particular cases $\alpha=0$ (no pressure dependence) and
$\alpha=1,\,2$ (linear and quadratic scaling with the pressure).

\subsection{Terms with $\alpha=0$.}

Starting with the $\alpha=0$ case we have
\be
\gamma + \delta = 2\beta.
\ee
The resulting form of $\Lambda$ in geometric units is
\be
\Lambda \sim (r^2 \rho)^\beta\left ( \frac{m}{r} \right )^\gamma.
\label{da6}
\ee
Formally, this combination is of order $( m/r)^{\beta+\gamma}$.
Therefore, we can generate $N$-PN terms if $\beta+\gamma=N$. These are
of the form
\be
\Lambda_N (\beta) \sim (r^2 \rho)^\beta\left ( \frac{m}{r} \right )^{N-\beta},
\label{L0}
\ee
with $\beta= 0, \pm1, \pm 2, ...$. For instance, the first few 1PN and 2PN terms of this series are (we start from $\beta=-1$ for
reasons explained below):
\bear
&&   \Lambda_1 (-1) \sim \frac{m^2}{r^4 \rho}, ~ \Lambda_1 (0) \sim \frac{m}{r}, ~ \Lambda_1 (1) \sim r^2 \rho,
\\
\nonumber \\
&& \Lambda_2 (-1) \sim \frac{m^3}{r^5 \rho},~ \Lambda_2 (0) \sim \frac{m^2}{r^2},~  \Lambda_2 (1) \sim r \rho m.
\eear

\subsection{Terms with $\alpha=1$.}

The $\alpha=1$ group of terms can be obtained with the same procedure. We have
\be
\gamma + \delta = 2(1+\beta),
\ee
and this leads to terms of the form
\be
\Lambda \sim r^2 p  (r^2 \rho)^\beta \left ( \frac{m}{r} \right )^\gamma.
\label{da10}
\ee
Since $r^2 p$ is a 2PN term, the resulting $N$-PN combination should take the form:
\be
\Lambda_N (\beta) \sim r^2 p  (r^2 \rho)^\beta \left ( \frac{m}{r} \right )^{N-2-\beta}.
\label{L1}
\ee
The first few 1PN and 2PN terms generated from this expression are:
\bear
&& \Lambda_1 (-1) \sim \frac{p}{\rho},\,~ \Lambda_1 (0) \sim \frac{r^3 p}{m}, ~ \Lambda_1 (1) \sim \frac{r^6 \rho p}{m^2}, ~
\\
\nonumber \\
&& \Lambda_2 (-1) \sim \frac{p m}{\rho r}, ~ \Lambda_2 (0) \sim r^2 p, ~ \Lambda_2 (1) \sim \frac{r^5 \rho p}{m}.
\eear

\subsection{Terms with $\alpha=2$.}

Finally, we consider the $\alpha=2$ terms. The corresponding $\Lambda_N$ combination is,
\be
\Lambda_N (\beta) = (r^2 p)^2 (r^2 \rho)^\beta \left ( \frac{m}{r} \right )^{N-4-\beta},
\label{L2}
\ee
and from this we have:
\bear
&& \Lambda_1 (-1) \sim \frac{r^4 p^2}{m^2 \rho}, \quad \Lambda_1 (0) \sim \frac{r^7 p^2}{m^3},
\eear
\bear
&&\Lambda_2(-1) \sim \frac{r^3 p^2}{\rho m}, \quad \Lambda_2 (0) \sim \frac{r^6 p^2}{m^2}.
\eear


\subsection{Generic $N$-PN order terms and constraints.}

It is now not too difficult to see that a $N$-PN order term with an arbitrary $p^\alpha$ scaling and with $\Pi$ re-introduced
is given by the universal formula,
\be
\Lambda_N (\alpha, \beta,\theta) \sim \Pi^\theta (r^2 p)^\alpha  (r^2 \rho)^\beta \left ( \frac{m}{r} \right )^{N-2\alpha-\beta-\theta}.
\label{Lgen}
\ee
As discussed in Section~\ref{sec:pT2}, different threads of reasoning lead to the constraint $\beta \geq -1$
The first one has to do with avoiding a divergence at the stellar surface (this already has allowed us to filter out all
$\alpha<0$ terms).  An inspection of the two stellar structure equations reveals that terms with $\alpha=\theta=0$
should scale as
\be
\rho \Lambda_N (0,\beta,0) \sim \rho^{1+\beta},
\ee
in the vicinity of the surface, and therefore we ought to take
$\beta \geq -1$ in order to avoid a surface singularity.  This
argument still allows for $\beta<-1$ values in the $\Lambda_N $ terms
with $\alpha, \theta > 0$, since these terms have a smoother profile
as a result of the vanishing of $p$ and $\Pi$ at the surface.

The second thread is no more than a heuristic argument and has to do
with the expectation that for a broad family of gravity theories the
solution for the metric (and its derivatives) should scale as
$\sim (\epsilon + \tau p)^n = \rho^n ( 1 + \Pi + \tau p/\rho)^n$ with
the fluid parameters (where $\tau$ and $n$ are ${\cal O}(1)$
numbers). From this it follows that negative powers of $\rho$ will
come in the form of dimensionless PN terms
$\sim \rho^{n-1} (p/\rho)^k$, where $k=n,\,n-1,\dots$ (note that a
factor $\rho$ has been absorbed by the Newtonian prefactor in the
structure equations).  As a consequence, $\rho^{-1}$ is the only
possible negative power in a PN expansion.  Obviously, this argument
automatically takes care of the regularity of any
$\Lambda_N (\alpha,\beta,\theta)$ term at the surface.

The exclusion of all $\alpha \geq 4$ terms comes about as a consequence of regularity
at the stellar center. Near the origin (where $p,\rho,\Pi$ take finite
non-zero values) a $\Lambda_N(\alpha,\beta,\theta)$ term behaves as
\be
\Lambda_N (r\to 0) \sim  r^{2(N-\alpha -\theta)}.
\ee
The corresponding terms in the stellar structure equations will behave as
\bear
&& \frac{dp}{dr} \sim \frac{\rho m}{r^2} \Lambda_N \sim r^{2(N-\alpha -\theta) +1},
\label{dpdr_r0}
\\
\nonumber \\
&& \frac{dm}{dr} \sim r^2 \rho  \Lambda_N \sim r^{2(N-\alpha -\theta +1)}.
\label{dmdr_r0}
\eear
and therefore regularity at the center dictates the following limits for each equation:
\bear
&& \frac{dp}{dr}: \quad 0 \leq \alpha \leq N -\theta,
\label{a_lim1a}
\\
&& \frac{dm}{dr}: \quad 0 \leq \alpha \leq N+1-\theta.
\label{a_lim1b}
\eear
We can also see that these conditions entail the following limits for $\theta$:
\be
\frac{dp}{dr}: \quad 0 \leq \theta \leq N, \qquad \frac{dm}{dr}: \quad 0 \leq \theta \leq N + 1.
\ee
For the particular case of 2PN order terms we then have:
\bear
&& \frac{dp}{dr}: \quad  0 \leq \theta \leq 2, \quad 0 \leq \alpha \leq  2-\theta,
\\
&& \frac{dm}{dr}: \quad 0 \leq \theta \leq 3, \quad 0 \leq \alpha \leq 3-\theta,
\eear
which shows that all $\alpha\geq 4$ terms are to be excluded and that
$\alpha=3$ terms can only appear in the mass equation.


\section{The Newtonian and relativistic Lane-Emden equations}
\label{sec:RLE}

In this appendix we review the nonrelativistic and relativistic Lane-Emden equations. The former equation
is classic textbook material (see e.g.~\cite{Chandrasekhar}) and therefore is just sketched here. The somewhat less
familiar relativistic extension was developed by Tooper \cite{Tooper:64,Tooper:65} and is discussed in a bit more
detail below. Our definition for the polytropic EOS, i.e. $p = K\rho^{1+1/n}$, is the same as the one adopted in \cite{Tooper:65}
but is different to the one used in Tooper's earlier paper~\cite{Tooper:64}, i.e. $p =  K\epsilon^{1+1/n}$.
This subtle difference, combined with the choice between $p_c/\rho_c$ or $p_c/\epsilon_c$
(the ``c" index refers to the stellar center) for the scale of the system, leads to slightly different Lane-Emden equations.


\subsection{The Newtonian Lane-Emden equation}

In Newtonian gravity, one can express the hydrostatic equilibrium
equation for spherical non-rotating stars in terms of dimensionless
parameters for the pressure, the density and the radial coordinate. If
the EOS is polytropic (i.e., according to our definition,
$p=K\rho^{1+1/n}$) the equations governing the dimensionless
quantities are scale-invariant, depending only on the polytropic index
$n$. By writing the density and the pressure as
\be
 \theta^n \equiv \frac{\rho}{\rho_c},
 \quad
 p = K \rho_{\rm c}^{1+1/n} \theta^{n+1},
\ee
and introducing the dimensionless radial coordinate
\be
r = \alpha\xi, \qquad
\alpha \equiv \left[\frac{(n+1) K}{4\pi G} \rho_{\rm c}^{-1+1/n}\right]^{1/2},
\ee
the Newtonian stellar structure equations
\begin{align}
\frac{dp}{dr} &= -\frac{Gm_{\rm N}}{r^2} \rho, 
\label{dpN1}
\\
\frac{dm_{\rm N}}{dr} &= 4\pi r^2 \rho,
\label{dmN1}
\end{align}
lead to
\be
\label{LE:nr}
\frac{1}{\xi^2}\frac{d}{d\xi}\left(\xi^2\frac{d\theta}{d\xi}\right)=-\theta^n.
\ee
This is the famous Lane-Emden equation, and its scale-invariant solutions describe all possible fluid 
configurations in terms of the single parameter $n$.


\subsection{The relativistic Lane-Emden equations}

Generalizing the Lane-Emden formalism to GR is a straightforward task,
but this comes at the price of losing the scale-invariance property of
the Newtonian treatment. In relativity we can define the polytropic
EOS in the same way as before, where $\rho$ is the baryonic rest mass
density. The polytropic exponent is defined as
\be\label{gamaPoly}
\Gamma=1+\frac{1}{n}=\frac{\rho}{p} \frac{dp}{d\rho}=\frac{\epsilon+p}{p}\frac{dp}{d\epsilon}.
\ee
Then the energy density $\epsilon$ and the internal energy $\Pi$ are given by
\be
\epsilon=\rho+np,
\label{eq:poly}
\ee
which implies
\be
\Pi = n\frac{p}{\rho}.
\ee
This observation was used in the argument leading to Eq.~\eqref{Fequiv}.

We can now introduce the relativistic version of the Lane-Emden
equations. In analogy with the Newtonian case we define
$\rho = \rho_{\rm c}\theta^n$, $r = a\xi$, and
$p=K\rho_{\rm c}^{1+1/n}\theta^{n+1}$.
The ratio between the central pressure and the central energy density
\be
\lambda \equiv \frac{p_{\rm c}}{\epsilon_{\rm c}}=\frac{K\rho_{\rm c}^{1+1/n}}{\rho_{\rm c}+nK\rho_{\rm c}^{1+1/n}},
\label{lam_scale}
\ee
is a convenient measure of the importance of relativistic effects in the system.
Note that our definition deviates from Tooper's~\cite{Tooper:65},
who prefers to use the ratio $p_{\rm c}/\rho_{\rm c}$.

The energy density is then
\bear
&& \epsilon=\rho_{\rm c}\theta^n+nK\rho_{\rm c}^{1+1/n}\theta^{n+1}
\nonumber \\
&& ~ =\epsilon_{\rm c}\left[1+n\lambda(\theta-1)\right]\theta^n.
\eear

We now want to derive a dimensionless form of the TOV equations~\eqref{TOV}.
The definition of the mass function $m_{\rm T}$ implies
\be
\frac{dm_{\rm T}}{d\xi} = 4\pi \epsilon_{\rm c} a^3[1+n\lambda(\theta-1)]\theta^n\xi^2.
\ee
In terms of the dimensionless mass
\be
\bar{m} \equiv \frac{m_{\rm T}}{a^3 \epsilon_c},
\ee
this becomes
\be
\frac{d\bar{m}}{d\xi}= 4\pi[1+n\lambda(\theta-1)]\theta^n\xi^2.
\ee
From the TOV equation for the pressure we similarly obtain, after some manipulations,
\begin{align}
\frac{d\theta}{d\xi}&= -\frac{\bar{m}}{\xi^2}\left(1-n\lambda\right)\left[1+(n+1)\frac{\lambda}{1-n\lambda}\theta\right]\nonumber\\
&\times \left(1+\lambda\frac{4\pi\xi^3\theta^{n+1}}{\bar{m}}\right)\left[1-2(n+1)\lambda\frac{\bar{m}}{\xi}\right]^{-1}.
\end{align}
In the present case the characteristic length scale is
\be
a=\left[(n+1)K\rho_{\rm c}^{-1+1/n}\left(1-n\lambda\right)^2\right]^{1/2}.
\ee

At this point we would like to define dimensionless quantities that come from the relativistic Lane-Emden equations.
The central baryonic rest-mass density is related to $\lambda$ as [see Eq.~(\ref{lam_scale})]:
\be
\rho_{\rm c}= K^{-n} \ell^n, \quad \ell  \equiv \frac{\lambda}{1-n\lambda}.
\ee
The factor $K^{-n}$ has units of mass density (or inverse square
length in geometrical units), therefore the dimensionless rest-mass
density is
\be
\bar{\rho} \equiv \rho K^n =\ell^n\theta^n.
\ee
Similarly, the length scale $a$ takes the form
\be
a=K^{n/2}\sqrt{(n+1)\ell^{1-n}}\left(1-n\lambda\right),
\ee
where $K^{n/2}$ has dimensions of length. The dimensionless radius is defined as
\be
\bar{r} \equiv r K^{-n/2} =\sqrt{(n+1)\ell^{1-n}}\left(1-n\lambda\right) \xi.
\ee
The remaining dimensionless parameters are
\begin{align}
\bar{\epsilon}& \equiv \epsilon K^n  =\left(\frac{\ell^n}{1-n\lambda} \right)\left[1+n\lambda(\theta-1)\right]\theta^n,\\
\nonumber \\
\bar{\mu}& \equiv m_{\rm T}  K^{-n/2}  \\
& =\left[\sqrt{(n+1)\ell^{1-n}}\left(1-n\lambda\right)\right]^3 \left(\frac{\ell^n}{1-n\lambda} \right)\bar{m}, \\
\nonumber \\
\bar{p}& \equiv p K^n =\ell^{n+1}\theta^{n+1},\\
\nonumber \\
\Pi&=n \frac{\bar{p}}{\bar{\rho}} = n\,\ell \theta.
\end{align}
All of the above dimensionless profiles are functions of $\xi$, $n$
and $\lambda$.  At variance with the Newtonian treatment, the
relativistic Lane-Emden formalism does not allow for a simple algebraic
mass-radius relation $M(R)$. This is also related to the fact that the system is not
scale-invariant, due to the presence of $\lambda$ in the equations.


\section{The PPN Potentials}
\label{sec:potentials}

The goal of this appendix is to study the behavior of the potentials
$U$, $E$ and $\Omega$ appearing in the PPN stellar structure
equations~\eqref{PPN_CR}, first derived by Ciufolini and
Ruffini~\cite{CiufoliniRuffini:1983}. By means of a mass function
redefinition (see Section~\ref{sec:PPN}) these potentials can be
eliminated at 1PN order, but they could still appear at 2PN order and
higher.

Given the 2PN precision of our calculations we can write these potentials as:
\begin{subequations}
\label{eq:potentials}
\begin{align}
U(r) &= -\int_{0}^{r} dr^{\p} \,\frac{m_{\rm N}}{r^{\p 2}} + U(0),
\label{Usol1}
\\
E(r) &= 4 \pi \int_{0}^{r} dr^{\p} r^{\p 2} \rho \Pi,
\label{Esol1}
\\
\Omega(r) &= -4\pi \int_{0}^{r} dr^{\p} \, r^{\p} \rho \, m_{\rm N},
\label{Omsol1}
\end{align}
\end{subequations}
where all right-hand side quantities are computed in Newtonian theory.
In Eqs.~\eqref{eq:potentials}, $m_{\rm N}(r)$ denotes the Newtonian mass function
\begin{align}
m_{\rm N}(r) &= 4 \pi \int_{0}^{r} dr^{\p}  \, \rho r^{\p 2}
= 4 \pi m_{\rm b} \int_{0}^{r} dr^{\p}\, n_{\rm b} r^{\p 2},
\end{align}
where $n_{\rm b}$ is the baryon number density. The integral
quantities $U$, $E$ and $\Omega$ represent the system's gravitational
potential energy, internal energy, and gravitational potential energy
respectively~\cite{poisson2014gravity}.  They appear as dimensionless
PN terms in the form of reduced potentials: $U$, $E/m_{\rm N}$,
$\Omega/m_{\rm N}$ [see Eqs.~(\ref{PPN_CR})].

The radial profiles of the three potentials inside the star can be
determined by first integrating the Newtonian hydrostatic equilibrium
equations, Eqs.~(\ref{dpN1}) and (\ref{dmN1}),
%
%
to find $m_{\rm N}$ and $p$ as functions of $r$. Using realistic EOS
data tables for $p(\rho)$ we can subsequently compute the internal density per unit
mass $\Pi(p)$ and the mass density $\rho(p) = m_{\rm b} n_{\rm b}(p)$,
and then numerically evaluate the potentials inside the star by
integration.


\begin{figure}[t]
\includegraphics[width=0.49\textwidth]{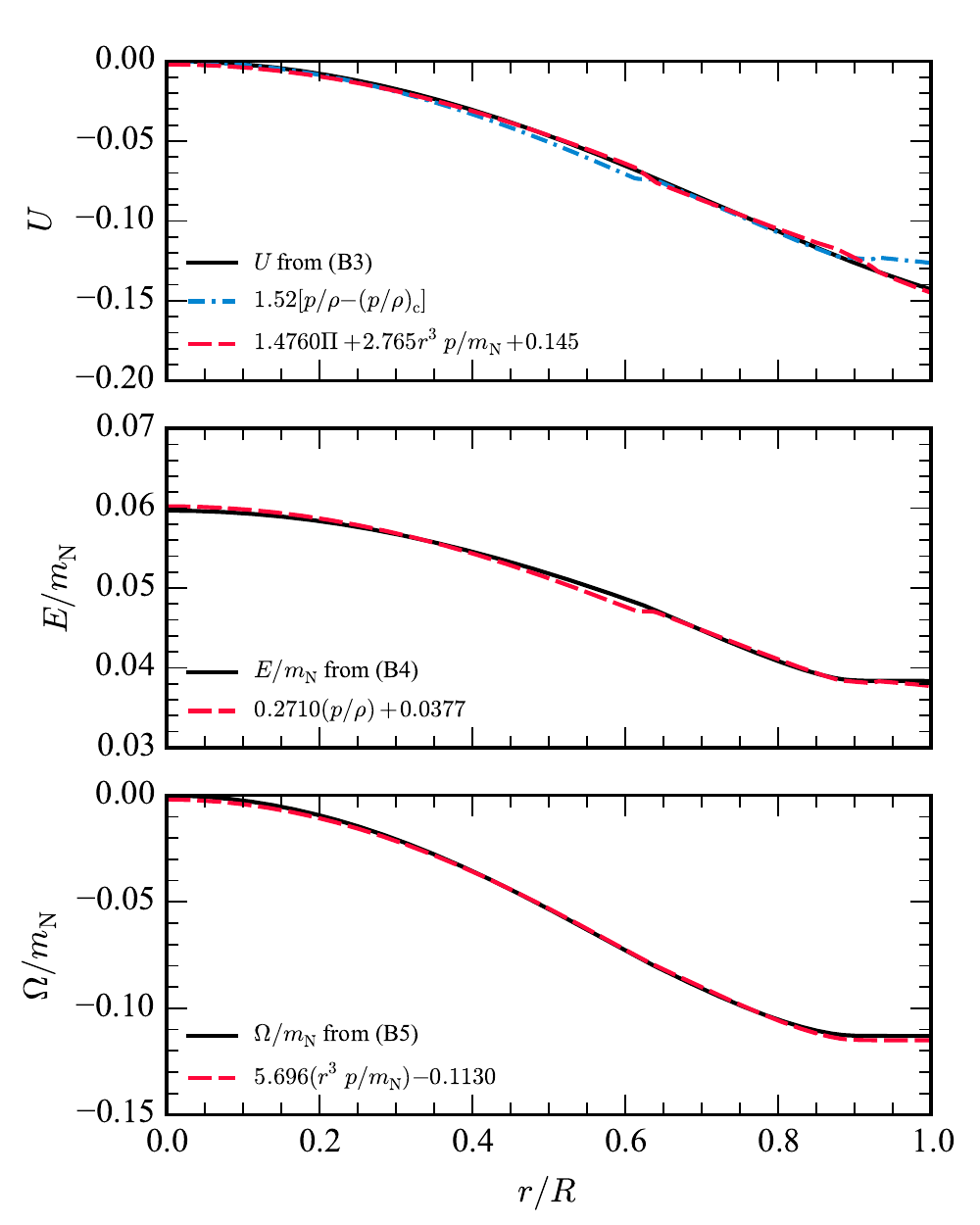}
\caption{{\it Integral PN potentials and 1PN terms.} The radial
  profiles of the integral potentials $U$, $E/m_{\rm N}$ and
  $\Omega/m_{\rm N}$ are well fitted by linear functions of the
  non-integral potentials $p/\rho$ and $r^3p/m_{\rm N}$. In all plots,
  the radial coordinate is normalized to the stellar radius $R$.}
\label{fig:e_over_m}
\end{figure}

Some insight into the nature of these potentials can be obtained by
rewriting Eqs.~(\ref{eq:potentials}) in the form
\begin{subequations}
\label{pot_solutions}
\begin{align}
\label{potentEquiv1}
U &= \frac{m_{\rm N}}{r} + 4\pi \int_r^R dr^\prime r^\p \rho,
\\
\label{potentEquiv2}
\frac{E}{m_{\rm N}} &= \Pi -\frac{1}{m_{\rm N}} \int_0^r dr^\prime\, m_{\rm N} \frac{d\Pi}{dr^\p},
\\
\frac{\Omega}{m_{\rm N}} &= -\frac{m_{\rm N}}{2r} -\frac{1}{2m_{\rm N}} \int_0^r dr^\p
\left( \frac{m_{\rm N}}{r^\prime} \right)^2
\nonumber
\\
&=4\pi\frac{r^3 p}{m_{\rm N}}-\frac{12\pi}{m_{\rm N}} \int_0^r dr^\prime\, r^{\p 2} \,p.
\label{potentEquiv4}
\end{align}
\end{subequations}
Note that the integration constant for $U$ has been fixed by requiring
$U(R) = M/R$ at the stellar surface, while those for
$E$ and $\Omega$ have been set to zero in order to have
regularity of $E/m_{\rm N}$ and $\Omega/m_{\rm N}$ at $r=0$.
The values of the potentials at the stellar center are:
\be
U(0) = 4\pi \int_0^R dr\, r \rho, \quad \frac{\Omega}{m_{\rm N}} (0) = 0, \quad \frac{E}{m_{\rm N}} (0) = \Pi_{\rm c}.
\ee

From Eqs.~(\ref{pot_solutions}) we can see that
$E/m_{\rm N}$ and $\Omega/m_{\rm N}$ are (partially) expressed in terms of the
non-integral 1PN terms
\be
\frac{m_{\rm N}}{r},\, \Pi,\,  \frac{r^3p}{m_{\rm N}} .
\label{3terms}
\ee
This suggests the possibility that the behavior of all three
potentials could be captured by linear combinations of non-integral
1PN terms. If true, this would mean that any 2PN term involving
$U, E/m_{\rm N}$ or $\Omega/m_{\rm N}$ is effectively accounted for by
the presence of the other terms in the post-TOV formulas. For
instance, this idea can be demonstrated for $U$ and for the special
case of a polytropic system. Starting from (\ref{Usol1}) and
expressing $m_{\rm N}$ in terms of $dp/dr$, after an
integration by parts and use of~\eqref{gamaPoly} we arrive at
\be
\label{prholin}
U=(n+1)\left(\frac{p}{\rho}-\frac{p_{\rm c}}{\rho_{\rm c}}\right) + U(0).
\ee
We know that for a polytrope $\Pi = n p/\rho$, which means
that we can also write
\be
U= \frac{(n+1)}{n} \left(\Pi- \Pi_{\rm c}\right) + U(0).
\ee
For a polytropic model, therefore, $U$ can be written exactly as a
linear function of $p/\rho$ or $\Pi$.

We have verified that $U$, $E/m_{\rm N}$ and $\Omega/m_{\rm N}$ can be
approximated by similar linear functions for the case of realistic
EOSs. As an illustration, in Fig.~\ref{fig:e_over_m} we consider a
stellar model built using the APR EOS with central mass density of
$0.58\times 10^{15}$ g/cm$^3$, Newtonian mass
$m_{\rm N} = 1.50 M_{\odot}$ and radius $R = 14.8$ km. For this model
we plot the radial profiles of $U$ (top panel), $E/m_{\rm N}$ (middle
panel) and $\Omega/m_{\rm N}$ (bottom panel).  The figure shows that
the profiles of the three potentials can be accurately reproduced by
linear combinations of the 1PN terms in Eq.~\eqref{3terms}, and that $U$
is reasonably well fit by a linear function of $p/\rho$, as suggested
by~\eqref{prholin}.  This latter fit breaks down near the surface, but
with a different combination of 1PN terms (namely, $\Pi$ and
$r^3 p/m_{\rm N}$) one can produce a near-perfect fit.

In conclusion, the addition of the integral potentials $U$,
$E/m_{\rm N}$ and $\Omega/m_{\rm N}$ in the 2PN terms is unnecessary
because their behavior can be captured by linear combinations of the
non-integral PN terms which are already included in the post-TOV
equations (\ref{PTOV_2PN_intro}).



%

\end{document}